\documentclass{aa}
\usepackage{graphicx}
\usepackage{txfonts}
\begin{document} 

   \title{Dynamically important magnetic fields near the event horizon of Sgr A*}

    \author{GRAVITY Collaboration\thanks{GRAVITY is developed
    in a collaboration between the Max Planck Institute for
    extraterrestrial Physics, LESIA of Observatoire de Paris/Universit\'e PSL/CNRS/Sorbonne Universit\'e/Universit\'e de Paris and IPAG of Universit\'e Grenoble Alpes /
    CNRS, the Max Planck Institute for Astronomy, the University of
    Cologne, the CENTRA - Centro de Astrofisica e Gravita\c c\~ao, and
    the European Southern Observatory. 
    \newline Corresponding author: A. Jim\'enez-Rosales (ajimenez@mpe.mpg.de)
    }:
A.~Jim\'enez-Rosales\inst{1}
\and J.~Dexter\inst{14,1}
\and F.~Widmann\inst{1}
\and M.~Baub\"ock\inst{1}
\and R.~Abuter\inst{8}
\and A.~Amorim\inst{6,13}
\and J.P.~Berger\inst{5}
\and H.~Bonnet\inst{8}
\and W.~Brandner\inst{3}
\and Y.~Cl\'{e}net\inst{2}
\and P.T.~de~Zeeuw\inst{11,1}
\and A.~Eckart\inst{4,10}
\and F.~Eisenhauer\inst{1}
\and N.M.~F\"orster~Schreiber\inst{1} 
\and P.~Garcia\inst{7,13}
\and F.~Gao\inst{1}
\and E.~Gendron\inst{2}
\and R.~Genzel\inst{1,12}
\and S.~Gillessen\inst{1}
\and M.~Habibi\inst{1}
\and X.~Haubois\inst{9}
\and G.~Heissel\inst{2}
\and T.~Henning\inst{3}
\and S.~Hippler\inst{3}
\and M.~Horrobin\inst{4}
\and L.~Jochum\inst{9}
\and L.~Jocou\inst{5}
\and A.~Kaufer\inst{9}
\and P.~Kervella\inst{2}
\and S.~Lacour\inst{2}
\and V.~Lapeyr\`ere\inst{2}
\and J.-B.~Le~Bouquin\inst{5}
\and P.~L\'ena\inst{2}
\and M.~Nowak\inst{17,2}
\and T.~Ott\inst{1}
\and T.~Paumard\inst{2}
\and K.~Perraut\inst{5}
\and G.~Perrin\inst{2}
\and O.~Pfuhl\inst{8,1}
\and G.~Rodr\'iguez-Coira\inst{2}
\and J.~Shangguan\inst{1}
\and S.~Scheithauer\inst{3}
\and J.~Stadler\inst{1}
\and O.~Straub\inst{1}
\and C.~Straubmeier\inst{4}
\and E.~Sturm\inst{1}
\and L.J.~Tacconi\inst{1}
\and F.~Vincent\inst{2}
\and S.~von~Fellenberg\inst{1}
\and I.~Waisberg\inst{15,1}
\and E.~Wieprecht\inst{1}
\and E.~Wiezorrek\inst{1} 
\and J.~Woillez\inst{8}
\and S.~Yazici\inst{1,4}
\and G.~Zins\inst{9}
}

\institute{
Max Planck Institute for Extraterrestrial Physics,
Giessenbachstra{\ss}e~1, 85748 Garching, Germany
\and LESIA, Observatoire de Paris, Universit\'e PSL, CNRS, Sorbonne Universit\'e, Universit\'e de Paris, 5 place Jules Janssen, 92195 Meudon, France
\and Max Planck Institute for Astronomy, K\"onigstuhl 17, 
69117 Heidelberg, Germany
\and $1^{\rm st}$ Institute of Physics, University of Cologne,
Z\"ulpicher Stra{\ss}e 77, 50937 Cologne, Germany
\and Univ. Grenoble Alpes, CNRS, IPAG, 38000 Grenoble, France
\and Universidade de Lisboa - Faculdade de Ci\^encias, Campo Grande,
1749-016 Lisboa, Portugal 
\and Faculdade de Engenharia, Universidade do Porto, rua Dr. Roberto
Frias, 4200-465 Porto, Portugal 
\and European Southern Observatory, Karl-Schwarzschild-Stra{\ss}e 2, 85748
Garching, Germany
\and European Southern Observatory, Casilla 19001, Santiago 19, Chile
\and Max Planck Institute for Radio Astronomy, Auf dem H\"ugel 69, 53121
Bonn, Germany
\and Sterrewacht Leiden, Leiden University, Postbus 9513, 2300 RA
Leiden, The Netherlands
\and Departments of Physics and Astronomy, Le Conte Hall, University
of California, Berkeley, CA 94720, USA
\and CENTRA - Centro de Astrof\'{\i}sica e
Gravita\c c\~ao, IST, Universidade de Lisboa, 1049-001 Lisboa,
Portugal
\and Department of Astrophysical \& Planetary Sciences, JILA, Duane Physics Bldg., 2000 Colorado Ave, University of Colorado, Boulder, CO 80309, USA
\and Department of Particle Physics \& Astrophysics, Weizmann Institute of Science, Rehovot 76100, Israel
\and Institute of Astronomy, Madingley Road, Cambridge CB3 0HA, UK
}

   \titlerunning{Strong magnetic fields near Sgr A*}
    
  \abstract{
We study the time-variable linear polarisation of Sgr A* during a bright NIR flare observed with the GRAVITY instrument on July $28^{\rm }$, 2018. 
Motivated by the time evolution of both the observed astrometric and polarimetric signatures, we interpret the data in terms of the polarised emission of a compact region (`hotspot') orbiting a black hole in a fixed, background magnetic field geometry. We calculated a grid of general relativistic ray-tracing models, created mock observations by simulating the instrumental response, and compared predicted polarimetric quantities directly to the measurements.
We take into account an improved instrument calibration that now includes the instrument's response as a function of time, and we explore a variety of idealised magnetic field configurations. 
We find that the linear polarisation angle rotates during the flare, which is consistent with previous results. The hotspot model can explain the observed evolution of the linear polarisation. 
In order to match the astrometric period of this flare, the near horizon magnetic field is required to have a significant poloidal component, which is associated with strong and dynamically important fields. The observed linear polarisation fraction of $\simeq 30\%$ is smaller than the one predicted by our model ($\simeq 50\%$). The emission is likely beam depolarised, indicating that the flaring emission region resolves the magnetic field structure close to the black hole.
}

\date{}


   \maketitle
%

\section{Introduction}

There is overwhelming evidence that the Galactic Centre harbours a massive black hole, Sagittarius A* \citep[Sgr A*,][]{ghez2008,genzel2010} with a mass of 
$M\sim4\times 10^6\ \rm{M}_\odot$ 
as inferred from the orbit of star S2 
\citep{schoedel2002,ghez2008,genzel2010,gillessen2017,gravity2017,gravity2018,gravity2019,gravity2020_precesion,do2019}. 
Due to its close proximity, Sgr A* has the largest angular size of any existing black hole that is observable from Earth, 
and it provides a unique laboratory to investigate the physical conditions of the matter and the spacetime around the object. 

The observed emission from Sgr A* is variable at all wavelengths from the radio to X-rays 
\citep[e.g.][]{baganoff2001,zhao2001,genzel2003,ghez2004,eisenhauer2005,macquart2006,marrone2008,eckart2008,do2009,witzel2018,do2019flare}. 
The simultaneous, large amplitude variations (`flares') seen in the near-infrared (NIR) and X-ray \citep{yusefzadeh2006,eckart2008sim} are the result of transiently heated relativistic electrons near the black hole, which are likely heated in shocks or by magnetic reconnection \citep{markoff2001,yuan2003,barriere2014,haggard2019}. 

The linear polarisation fraction of $\simeq 10-40\%$ \citep{eckart2006,trippe2007,eckart2008,zamaninasab2010,shahzamanian2015} implies that the NIR emission is the result of synchrotron emission from relativistic electrons. The NIR to X-ray spectral shape favours direct synchrotron radiation from electrons up to high energies \citep[$\gamma \sim 10^5$,][]{doddseden2009,li2015,ponti2017}, although inverse Compton scenarios may remain viable \citep{porquet2003,eckart2010,yusefzadeh2012}.

Using precision astrometry with the second generation beam combiner instrument GRAVITY at the Very Large Telescope Interferometer (VLTI) operating in the NIR \citep{gravity2017}, we recently discovered continuous clockwise motion that is associated with three bright flares from Sgr A* \citep{gravity2018flare,gravity2020_michi}. The scale of the apparent motion $\simeq 30-50 \ \mu$as is consistent with compact orbiting emission regions \citep[`hotspots', e.g. ][]{broderickloeb2005,broderickloeb2006,hamaus2009} at $\simeq 3-5 R_S$, where $R_S = 2GM/c^2 \simeq 10 \ \mu\rm{as}$, is the Schwarzschild radius. 
In each flare, we also find evidence for a continuous rotation of the linear polarisation angle. 
The period of the polarisation angle rotation matches what is inferred from astrometry. An orbiting hotspot sampling a background magnetic field can explain the polarisation angle rotation, as long as the magnetic field configuration contains a significant poloidal component. For a rotating, magnetised fluid, remaining poloidal in the presence of orbital shear implies a dynamically important magnetic field in the flare emission region.

Here, we analyse the GRAVITY flare polarisation data in more detail, accounting for an improved instrument calibration that now includes the VLTI's response as a function of time (Section \ref{sec:polcal}).
We find general agreement with our previous results of an intrinsic rotation of the polarisation angle during the flare
by using numerical ray tracing simulations (Section \ref{sec:model}); we created mock observations by folding hotspot models forward through the observing process. We compare this directly to the data to show that the hotspot model can explain the observed polarisation evolution as well as to constrain the underlying magnetic field geometry and viewer's inclination (Section \ref{sec:fitting}). 
Matching the observed astrometric period and linear polarisation fraction requires a significant poloidal component of the magnetic field structure on horizon scales around the black hole as well as an emission size that is big enough to resolve it. We discuss the  implications of our results and limitations of the simple model in Section \ref{sec:discussion}.


\section{GRAVITY Sgr A* flare polarimetry}
\label{sec:polcal}

   \begin{figure*}
   \centering
   \includegraphics[width=0.35\textwidth]{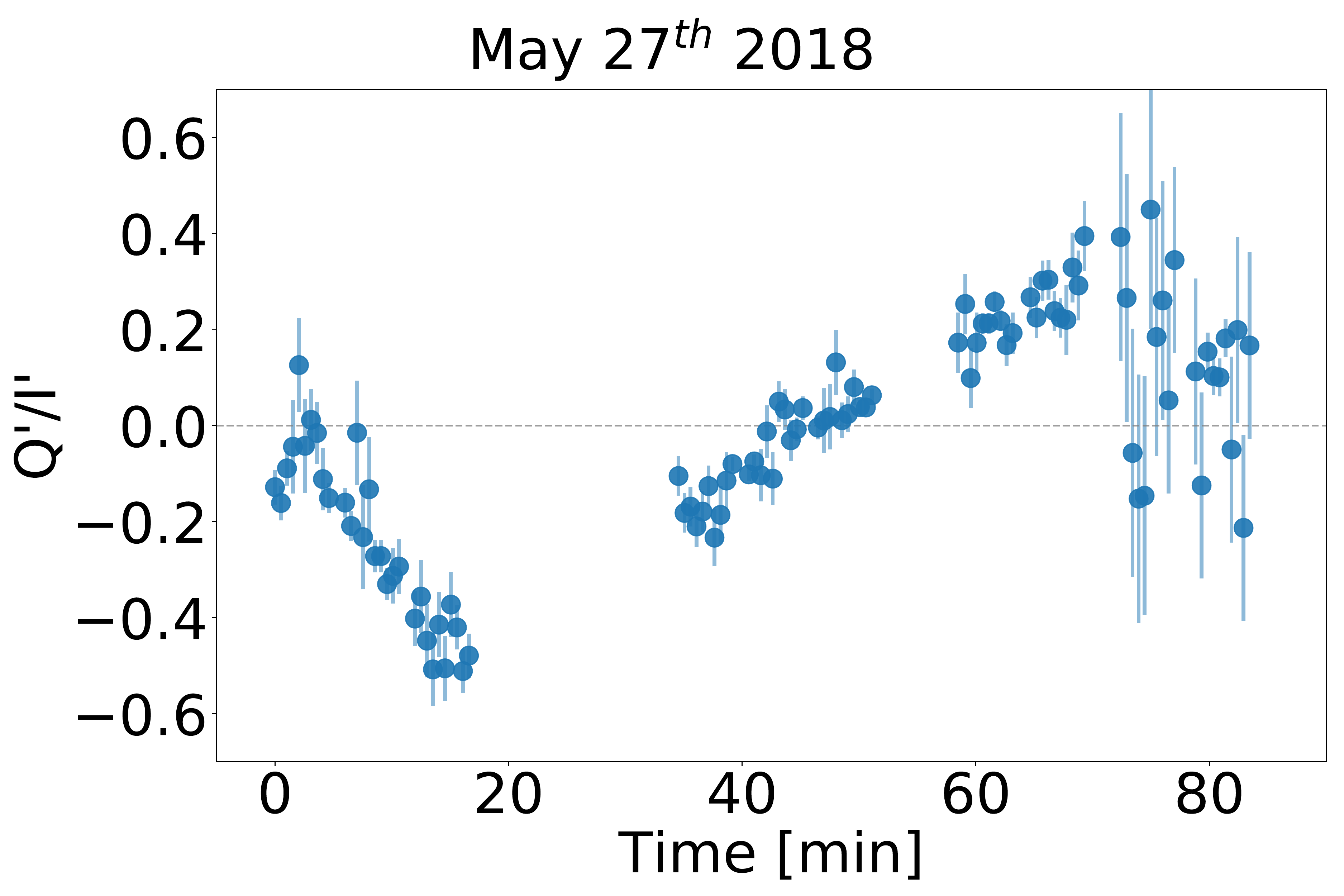}
    \includegraphics[width=0.355\textwidth]{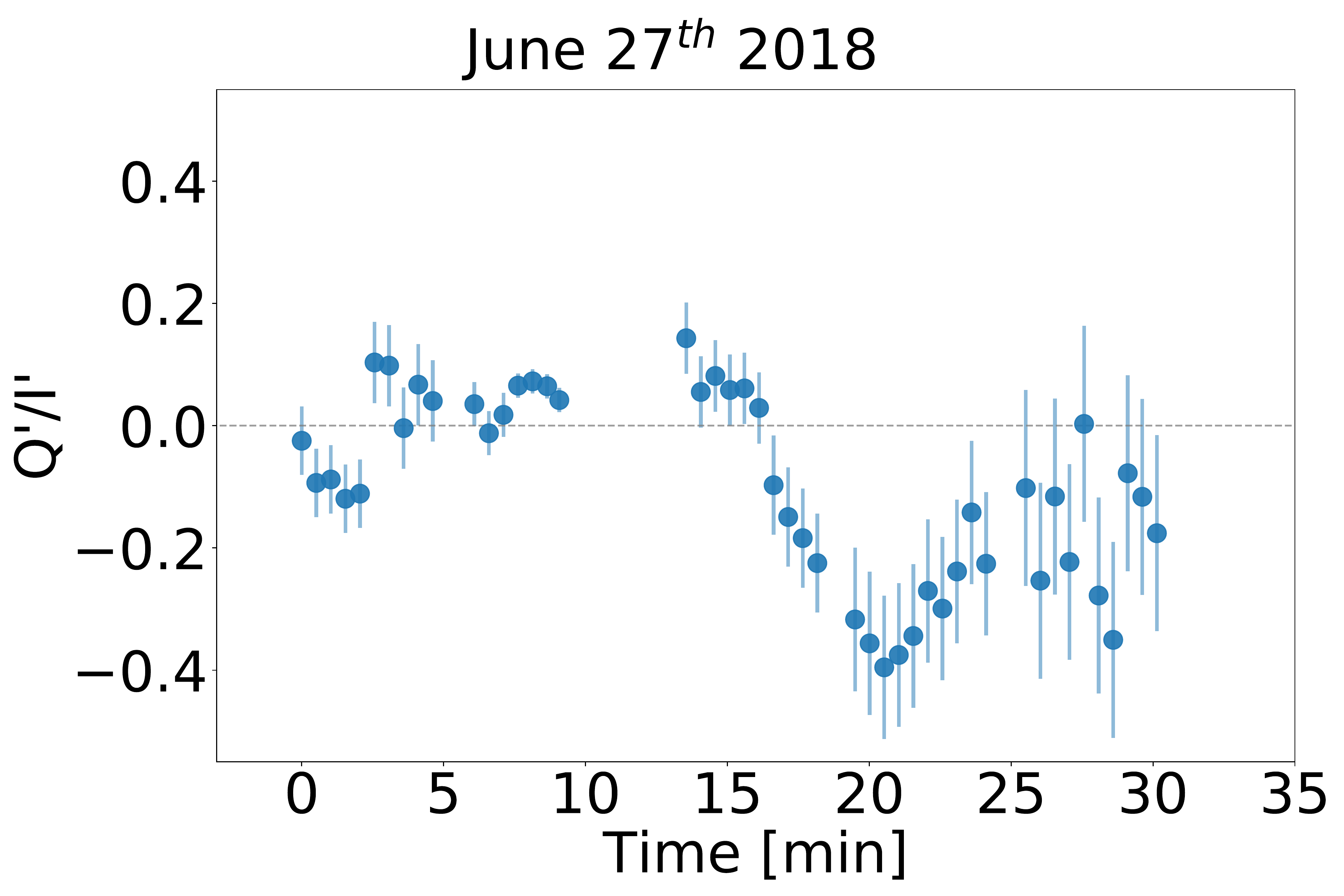}
    \includegraphics[width=0.35\textwidth]{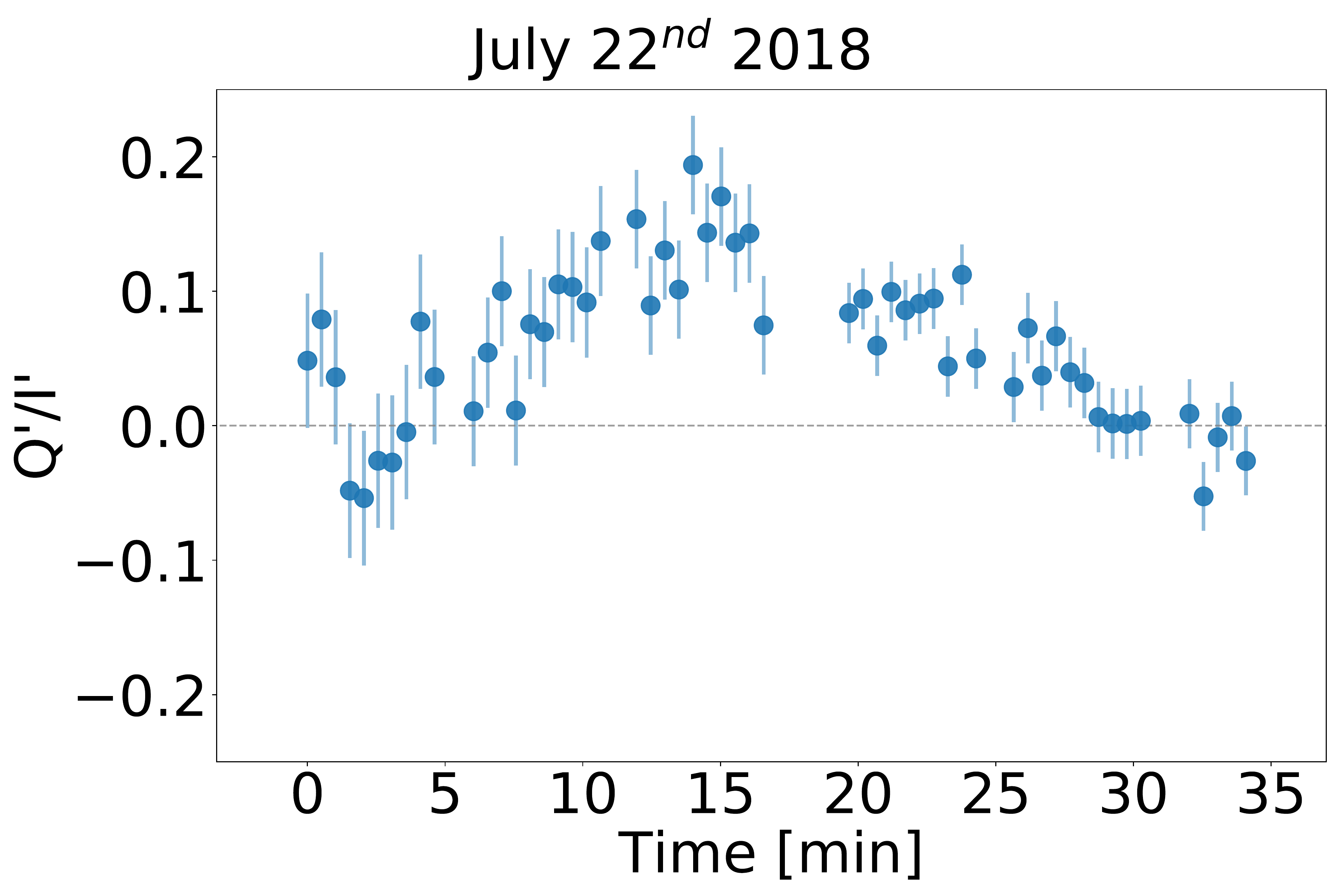}
   \includegraphics[width=0.35\textwidth]{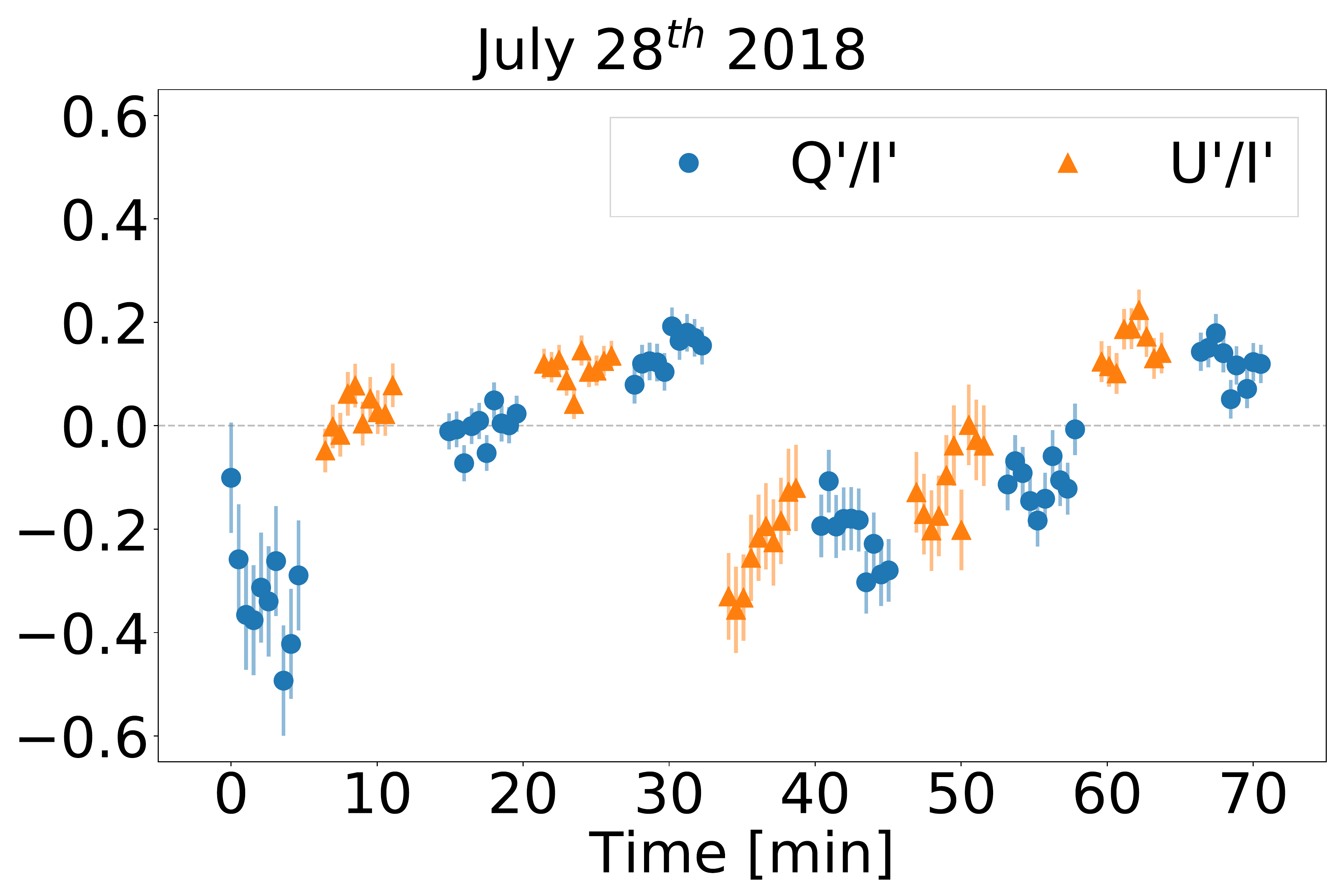}
   \caption{Linear polarisation Stokes parameters of four Sgr A* NIR flares observed by GRAVITY during 2018. The prime notation denotes the quantities as recorded by the instrument (including the effects of field rotation and systematics). Top left: Stokes $Q'$ on May $27^{\rm }$. Top right: Stokes $Q'$ on June $27^{\rm }$. Bottom left: Stokes $Q'$ on July $22^{\rm }$. Bottom right: Stokes $Q'$ and $U'$ on July $28^{\rm }$. All of the flares show $\gtrsim 10-40\%$ linear polarisation. A common, continuous evolution is seen on all nights. In three cases, $Q'$ shows a change in sign, consistent with rotation of the polarisation angle. The implied period of the polarisation evolution matches what is seen in astrometry.}
   \label{fig:pol_lightcurves}%
    \end{figure*}
%
   \begin{figure*}
   \centering
   \includegraphics[trim = 0cm 0cm 0cm 0cm, clip=true,width=0.75\textwidth]{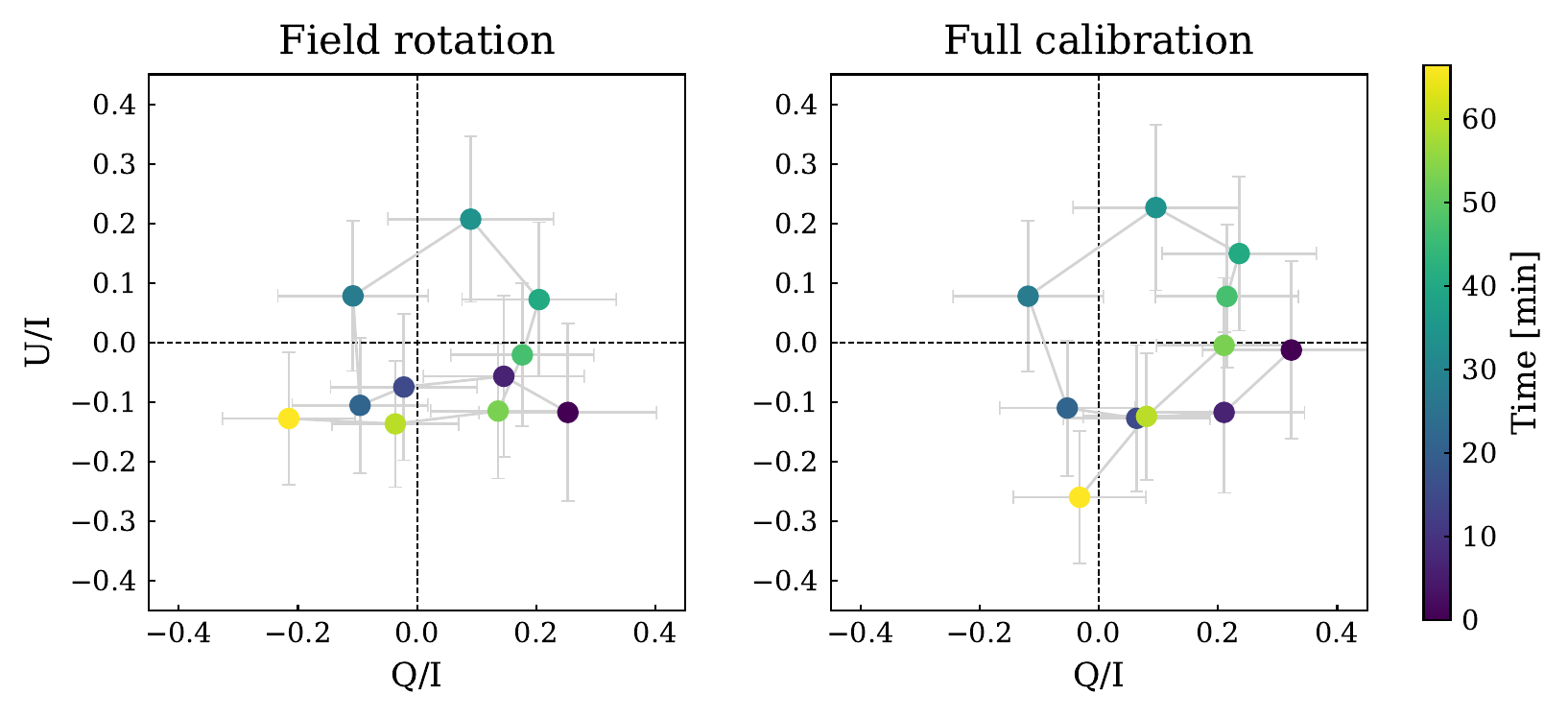}
   \caption{Reconstructed evolution of the on-sky linear Stokes parameters in $QU$ space for the July $28^{\rm }$ flare, linearly interpolating to fill in $U'$ and $Q'$ where the other is measured. Colour indicates time in minutes. Left: previous calibration where the quantities have only been subjected to a field rotation correction \citep{gravity2018flare}. Right: full new calibration including VLTI systematics and Stokes $V'$ reconstruction.
   In both cases, the flare traces 1.5 loops during its $60-70$ minute evolution.}
   \label{fig:july28_QUloop}%
    \end{figure*}
GRAVITY observations of Sgr A* have been carried out in split-polarisation mode, where interferometric visibilities are simultaneously measured in two separate orthogonal linear polarisations. A rotating half-wave plate can be used to alternate between the linear polarisation directions $P_{00}$ --- $P_{90}$ and $P_{-45}$ --- $P_{45}$.  
As a function of these polarised feeds, the Stokes parameters, as measured by GRAVITY, are $I'=(P_{00}+P_{90})/2$, $Q'=(P_{00}-P_{90})/2$ and $U'=(P_{45}-P_{-45})/2$. The circularly polarised component $V'$ cannot be recorded with GRAVITY. 

We relate on-sky (unprimed) polarised quantities with their GRAVITY measured (primed) counterparts by
\begin{equation}
\bar{S} = M \ \bar{S'}
\label{eq:MFT}
,\end{equation}
where $\bar{S}$ and $\bar{S'}$ are the on-sky and GRAVITY Stokes vectors, respectively, and $M$ is a matrix that characterises the VLTI's optical beam train response as a function of time, taking into account the rotation of the field of view during the course of the observations and birefringence. 
The former was calculated from the varying position of the telescopes during the observations and calibrated on sky by observing stars in the Galactic Centre \citep{gravity2018flare}. 
The latter are newly introduced in the analysis here and they were obtained from modelling the effects of reflections on a long optical path through the individual UT telescopes and the VLTI.

During 2018, GRAVITY observed several NIR flares from Sgr A* \citep{gravity2018flare}. Figure  \ref{fig:pol_lightcurves} shows the linear polarisation Stokes parameters for four of them as measured by the instrument.
On the top left, top right, and bottom left, the flares on May $27^{\rm }$, June $27,^{\rm }$ and July $22^{\rm }$ are shown, respectively. Only Stokes $Q'$ was measured on those nights. 
For the July $28^{\rm }$ flare (bottom right), both $Q'$ and $U'$ were measured. 
All of the flares observed during 2018 exhibit a change in the sign of the Stokes parameters during the flare, which is consistent with a rotation of the polarisation angle with time. 
The linear polarisation fractions are $\gtrsim 10-40\%$, which is in agreement with past measurements \citep{eckart2006,trippe2007,eckart2008}. 
Polarisation angle swings have also been previously seen in NIR flares with NACO \citep[e.g.][]{zamaninasab2010}. 
The smooth polarisation swings in both flares and the July $28^{\rm }$ single loop in $U$ versus $Q$ \citep[`$QU$ loop', ][Figure  \ref{fig:july28_QUloop}]{marrone2006} support the astrometric result of orbital motion of a hotspot close to event horizon scales of Sgr A*. 

Two assumptions have been made in the calculation of this loop. Since GRAVITY cannot register both linear Stokes parameters simultaneously, one has to interpolate the value of one quantity while the other is measured. In the case of Figure  \ref{fig:july28_QUloop}, this has been done by linearly interpolating between the median values over each exposure of $\simeq 5$ min. 
Second, no circular polarisation data are recorded (Stokes $V'$). This implies that transforming the GRAVITY measured Stokes parameters (primed) to on-sky values (unprimed) not only requires a careful calibration of the instrument systematics (contained in the matrix $M$, Eq. \ref{eq:MFT}), but an assumption on Stokes $V'$. In Figure  \ref{fig:july28_QUloop}, the assumption is that $V'=0$. 
While in theoretical models Stokes $V = 0$ is well justified for synchrotron radiation from highly relativistic electrons, birefrigence in the VLTI introduces a non-zero $V'$. It is therefore important to characterize it properly.

In this work, we adopt a forward modelling approach. We take intrinsic Stokes parameters $Q$ and $U$ from numerical calculations of a hotspot orbiting a black hole in a given magnetic field geometry, transform them to the GRAVITY observables $Q'$ and $U'$ following Eq. (\ref{eq:MFT}), and compare them to the data. This not only allows us to fit the July $28^{\rm{}}$ polarisation data directly without having to make assumptions on Stokes $V'$ or interpolate between gaps of data due to the lack of simultaneous measurements of the Stokes parameters, but to make predictions for $Q'$ when it is the only quantity measured, as is the case for the other 2018 flares.

\section{Polarised synchrotron radiation in orbiting hotspot models}
\label{sec:model}

   \begin{figure*}
   \centering
   \includegraphics[trim = 0cm 4cm 0cm 0cm, clip=true,width=0.7\textwidth]{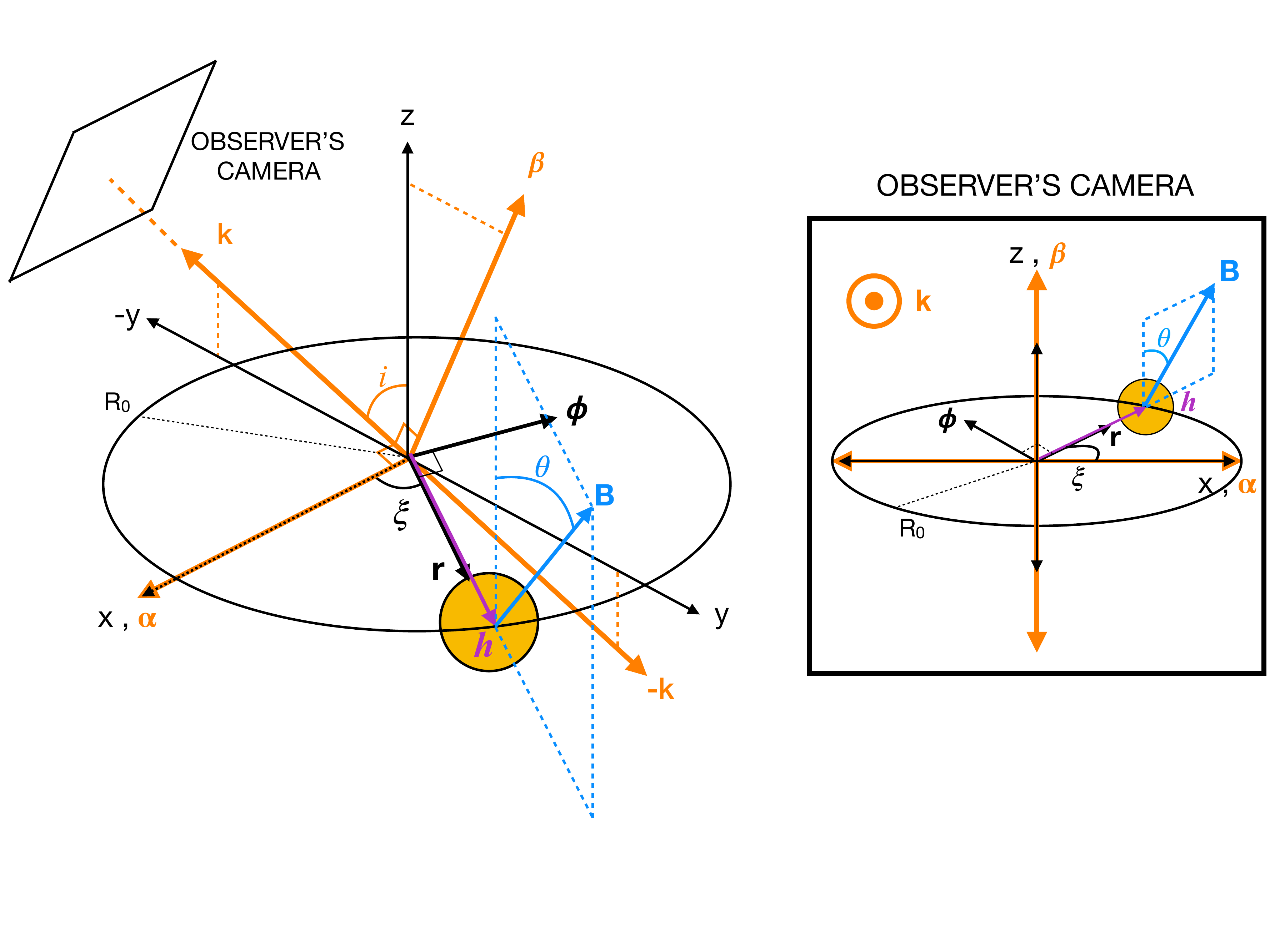}
   \caption{Lab frame diagram of a hotspot orbiting in the $\hat{x}\hat{y}$ plane with position vector $\bar{h} = R_0 \ \hat{r}$, where $\hat{r}$ is the unit vector in the radial direction. We note that $\bar{h}$ makes an angle $\xi(t)$ with $\hat{x}$. The magnetic field $\bar{B}$ is a function of $\xi$ and consists of a vertical plus radial component. The strength of the latter is given by $\tan{\theta}$, $\theta,$ the angle between the vertical, and $\bar{B}$. The observer's camera is defined by impact parameters $\hat{\alpha}$, $\hat{\beta}$, and a flat space line of sight $\hat{k}$. The line of sight makes an angle $i$ with the spin axis of the black hole. The observer's view is shown on the right. Lastly, $\hat{\phi}$ is the unit vector in the azimuthal direction.}
   \label{fig:hotspot_diagram}%
    \end{figure*}

   \begin{figure*}
   \centering
   \includegraphics[trim = 0 0 0cm 0, clip=false,width=0.60\textwidth]{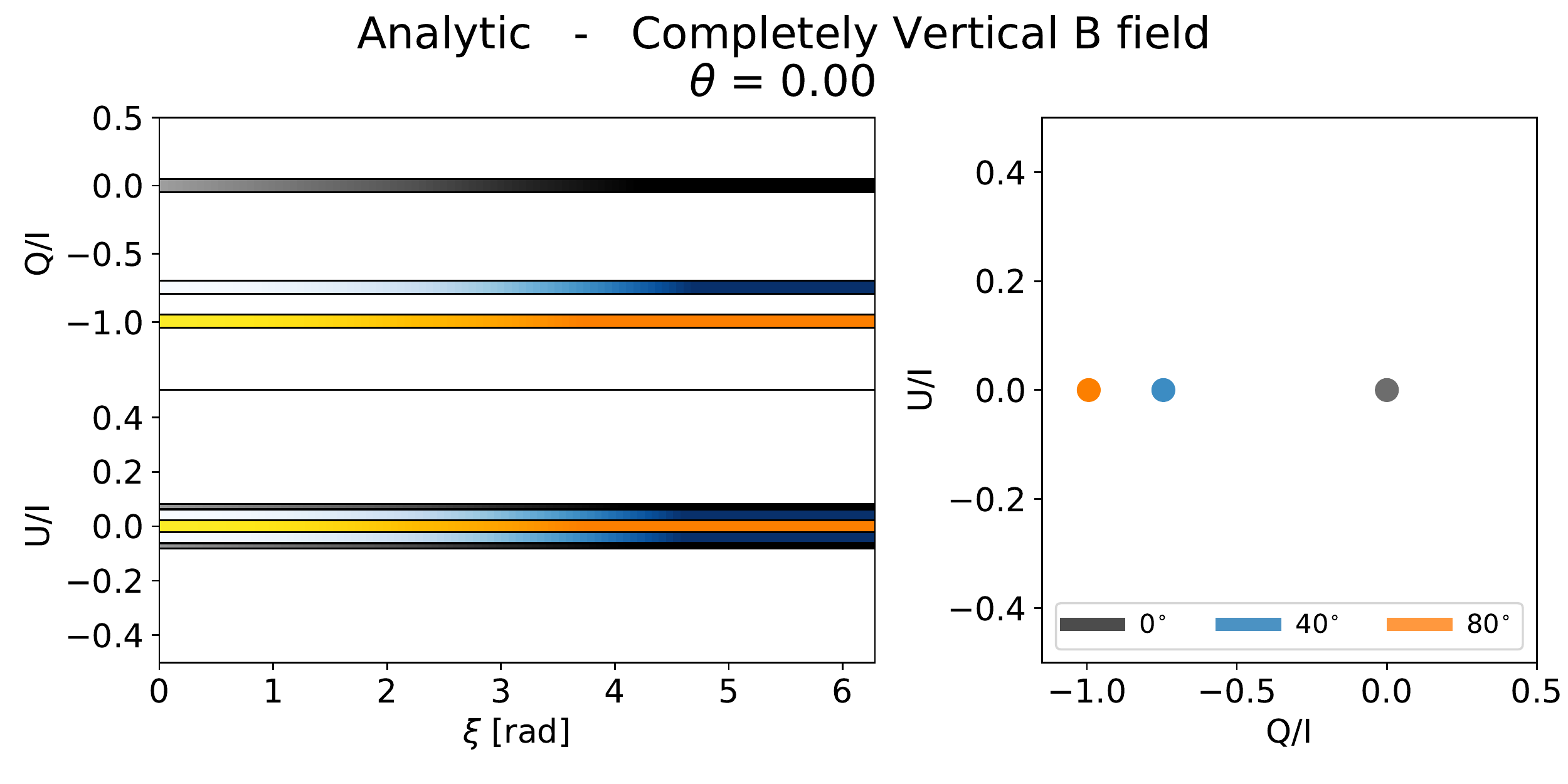}
   \includegraphics[trim = 0 0 0cm 0, clip=false,width=0.60\textwidth]{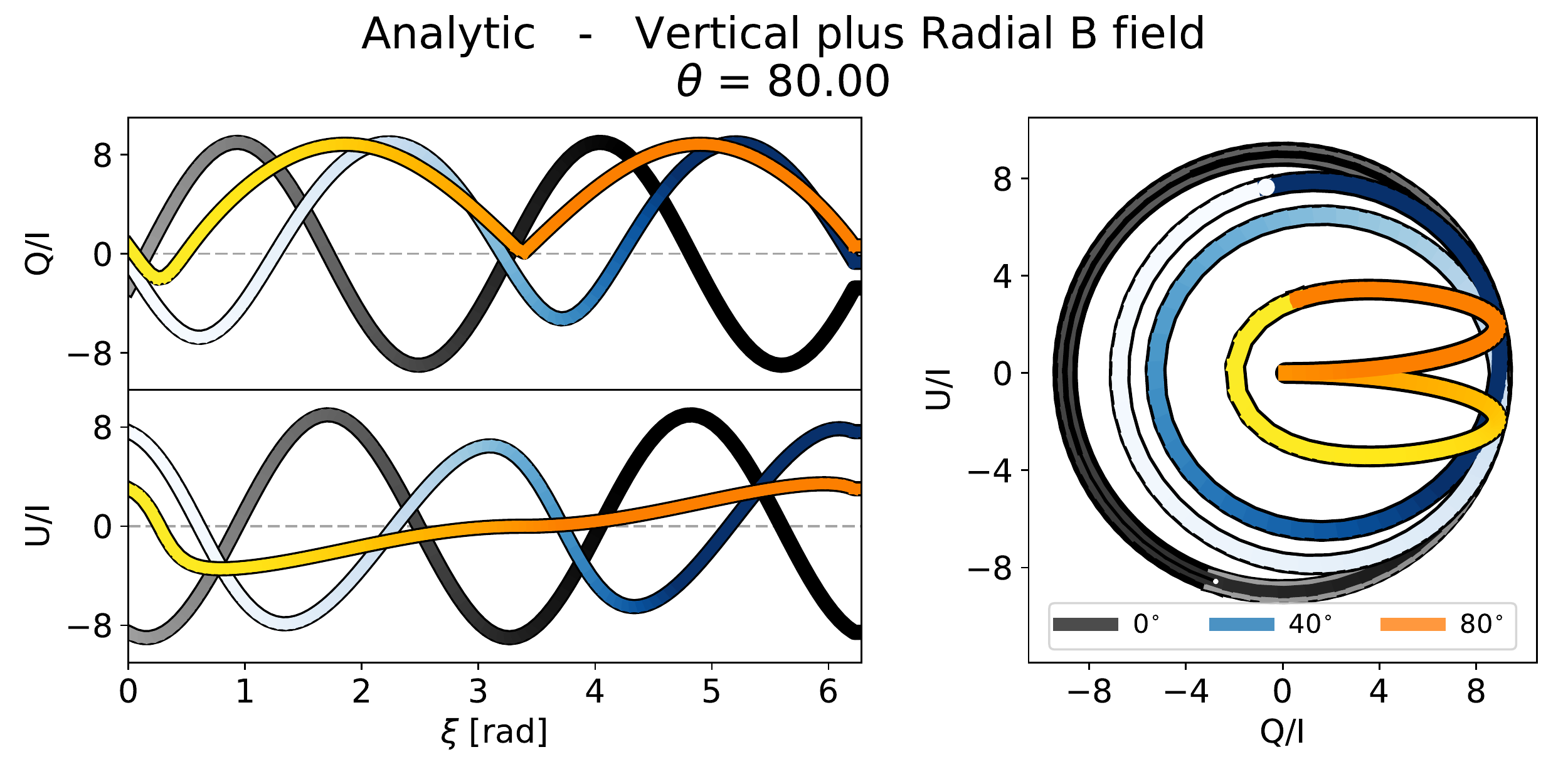}
   \caption{Analytic non-relativistic calculations of the linear Stokes parameters $Q$ and $U$ in a vertical plus radial magnetic field at three different viewer inclinations: $i=0^\circ,40^\circ,80^\circ$. The colour gradient denotes the periodic evolution of the hotspot along its orbit over one revolution. The only reason the width of the curves vary is for visualisation purposes.\  Top: completely vertical magnetic field ($\theta=0$). We note that $Q$ and $U$ are constants in time and have static values in $QU$ space. Bottom: significantly radial magnetic field with $\theta=80$; $Q$ and $U$ oscillate and trace two $QU$ loops in time that change in amplitude with inclination.  
   High inclination counteracts the presence of $QU$ loops. }
   \label{fig:analytic_examples}%
   \end{figure*}

An optically thin hotspot orbiting a black hole produces time-variable polarised emission, depending on the spatial structure of the polarisation map \citep{connors1977}. For the case of synchrotron radiation, the polarisation traces the underlying magnetic field geometry \citep{broderickloeb2005}. 
We first discuss an analytic approximation to demonstrate the polarisation signatures generated by a hotspot in simplified magnetic field configurations, before describing the full numerical calculation of polarisation maps used for comparison to the data.

\subsection{Analytic approximation}

We define the observer's camera centred on the black hole with impact parameters $\hat{\alpha}$ and $\hat{\beta}$, which are perpendicular and parallel to the spin axis, with a line of sight direction $\hat{k}$ \citep{bardeen1973}. In terms of these directions and assuming flat space, the Cartesian coordinates are expressed by 

\begin{equation}
\hat{x} = \hat{\alpha},\hspace{5pt}\\
\hat{y} = \cos{i} \ \hat{\beta} - \sin i \ \hat{k},\hspace{5pt}\\
\hat{z} = \sin{i} \ \hat{\beta} + \cos{i} \ \hat{k},
\end{equation}

\noindent where $i$ is the inclination of the spin axis to the line of sight. Equivalently,
\begin{equation}
\hat{\alpha} = \hat{x},\hspace{5pt}\\
\hat{\beta} = \cos{i} \ \hat{y} + \sin i \ \hat{z},\hspace{5pt}\\
\hat{k} = -\sin{i} \ \hat{y} + \cos{i} \ \hat{z}.
\end{equation}

\noindent When face-on, $\hat{k}$ points along $\hat{z}$ and $\hat{\beta}$ points along $\hat{y}$. When edge-on, $\hat{k}$ points along $-\hat{y}$ and $\hat{\beta}$ points along $\hat{z}$. 

Let a hotspot be orbiting in the $\hat{x}\hat{y}$ plane (Figure  \ref{fig:hotspot_diagram}). In terms of $\hat{\alpha}$, $\hat{\beta}$, and $\hat{k}$, the hotspot's position vector $\bar{h}$ is given by
\begin{align}
\bar{h} &= R_0 \ \hat{r} \nonumber \\
         &= R_0 \ ( \cos{\xi}\ \hat{\alpha}+\cos{i}\sin{\xi}\ \hat{\beta} -\sin{i}\sin{\xi}\ \hat{k} ),
\label{eq:r_vector}
\end{align}

\noindent where $\hat{r}$ is the canonical radial vector, $R_0$ is the orbital radius, and $\xi$ is the angle between $\hat{\alpha}$ and $\hat{r}$.

Let us consider the magnetic field with vertical and radial components given by

\begin{equation}
 \bar{B} \ = \frac{\ B_0 }{\sqrt{1+\delta^2}}\ (\hat{z}+\delta\ \hat{r})
         \hspace{5pt} ; \hspace{5pt} 
         \delta \equiv \tan{\theta}
    \label{eq:b_ver_analytic}
 ,\end{equation}
 
\noindent where $B_0$ is the magnitude of $\bar{B}$ and $\theta$ is the angle between $\hat{z}$ and $\bar{B}$.  
The polarisation is given as $\bar{P}=\hat{k} \times \bar{B}$. In flat space and in the absence of motion (no light bending or aberration),
\begin{align}
\bar{P} & \propto \hat{k}\times(\hat{z} + \delta\ \hat{r}) \nonumber \\
        & \propto  \ -(\sin{i} + \tan{\theta} \cos{i}\sin{\xi})\ \hat{\alpha} + \tan{\theta}\cos{\xi}\ \hat{\beta} \ .
\label{eq:P_ver+rad}
\end{align}

The polarisation angle on the observer's camera is $\tan\psi = \bar{P}\cdot\hat{\beta}/\bar{P}\cdot\hat{\alpha}$, so that
\begin{equation}
\psi = \tan^{-1} \left( -\frac{ \tan{\theta}\cos{\xi}} {\sin{i}+\tan{\theta}\cos{i}\sin{\xi}} \right).
\label{eq:pol_angle}
\end{equation}

Given that $U/Q=1/2\tan{\psi}$, the Stokes parameters as a function of the polarisation angle are
\begin{equation}
Q = |\bar{P}| \cos 2 \psi,\hspace{5pt}\\
U = |\bar{P}| \sin 2 \psi. 
\label{eq:QU_analytic}
\end{equation}

\noindent With equations (\ref{eq:P_ver+rad}), (\ref{eq:pol_angle}), and (\ref{eq:QU_analytic}), Stokes $Q$ and $U$ are obtained. 

It is important to note that a single choice of $i$ and $\theta$   returns $Q$=$Q$$(\xi)$ and $U$=$U$$(\xi)$. Assuming a constant velocity along the orbit, the angle $\xi$ can be mapped linearly to a time value by setting the duration of the orbital period and an initial position where the $\xi=0$. 

Additionally, an inclination of $i=i_0<90^\circ$ and $i=180^\circ-i_0$  produces the same polarised curves but they are reversed in $\xi$ with respect to each other.
This is expected since, for an observer at $i=i_0$ and one at $i=180^\circ-i_0$, the hotspot samples the same magnetic field geometry, but they appear to be moving in opposite directions with respect to each other.  
This means that the relative order in which the peaks in $Q$ and $U$ appear are reversed between observers at $i=i_0$ and at $i=180^\circ-i_0$. 

Given that light bending has not been considered in this approximation, in a significantly vertical field ($\theta\simeq0$, top of Fig. \ref{fig:analytic_examples}), the polarisation remains constant in $\xi$ (and time) proportional to $-\sin{i}$. In $QU$ space, this means a static value as the hotspot goes around the black hole. A particular case of this is $\bar{P}\simeq 0$ at $i\simeq 0$, since $\hat{k}$ and $\bar{B}$ are parallel. 
As $\theta\xrightarrow{}\pi/2$, $\tan{\theta}\xrightarrow{}\infty$ (bottom of Fig. \ref{fig:analytic_examples}), and the magnetic field becomes radial.
In this case and at low inclinations, the polarisation configuration is toroidal ($\bar{P}\propto\hat{\phi}$, the azimuthal canonical vector, Eq. \ref{eq:phi_vector}). As the hotspot orbits the black hole, $Q$ and $U$  show oscillations of the same amplitude. In one revolution, two superimposed $QU$ loops can be traced. If the viewer's inclination increases, one of the loops decreases more in size than the other and eventually disappears at very high inclinations, leaving only one behind. 
Increasing inclination, therefore, counteracts the presence of $QU$ loops in an analytical model with a vertical plus radial magnetic field. It is noted that the normalised polarisation configurations of a completely radial magnetic field and a toroidal one are equivalent with just a phase offset of $90^\circ$ in $\xi$ (Eq. \ref{eq:P_toroidal} in Appendix \ref{appendix:toroidal_B}).

\subsection{Ray-tracing calculations}

Next, we use numerical calculations to include general relativistic effects.  
We used the general relativistic ray tracing code \texttt{grtrans} \citep{dexteragol2009,dexter2016} to calculate synchrotron radiation from orbiting hotspots in the Kerr metric. 

The hotspot model is taken from \citet{broderickloeb2006}, and it consists of a finite emission region orbiting in the equatorial plane at radius $R_0$. The orbital speed is constant for the entire emission region, and it matches that of a test particle motion at its centre. 
The maximum particle density $n_{\rm spot}\sim2\times10^7 \, \rm cm^{-3}$ falls off as a three-dimensional Gaussian with a characteristic size of $R_{\rm spot}$. 
The magnetic field has a vertical plus radial component\footnote{ See Appendix \ref{appendix:ver+rad_kerr} for details.}. Its strength is taken from an equipartition assumption, where we further assume a virial ion temperature of  $k T_i = (n_{\rm spot}/n_{\rm tot}) \ (m_p c^2/R), \ (n_{\rm spot}/n_{\rm tot})=5$, where $n_{\rm tot}$ is the total particle density in the hotspot. For the models considered here, a typical magnetic field strength in the emission region is $B \simeq 100\ \rm{G}$. 
We calculated synchrotron radiation from a power law distribution of electrons with a minimum Lorentz factor of $1.5\times10^3$ and considered a black hole with a spin of zero.\footnote{Given the scales at which the hotspot is orbiting, a change in the spin of the black hole does not alter the results significantly. See Appendix \ref{apendix:spin_comparison} for more details}. The model parameters for field strength, density, and minimum Lorentz factor were chosen as typical values for models of Sgr A* which can match the observed NIR flux. Other combinations are possible.

   \begin{figure*}
   \centering
   \includegraphics[trim = 0cm 0cm 0cm 0cm, clip=true,width=0.7\textwidth]{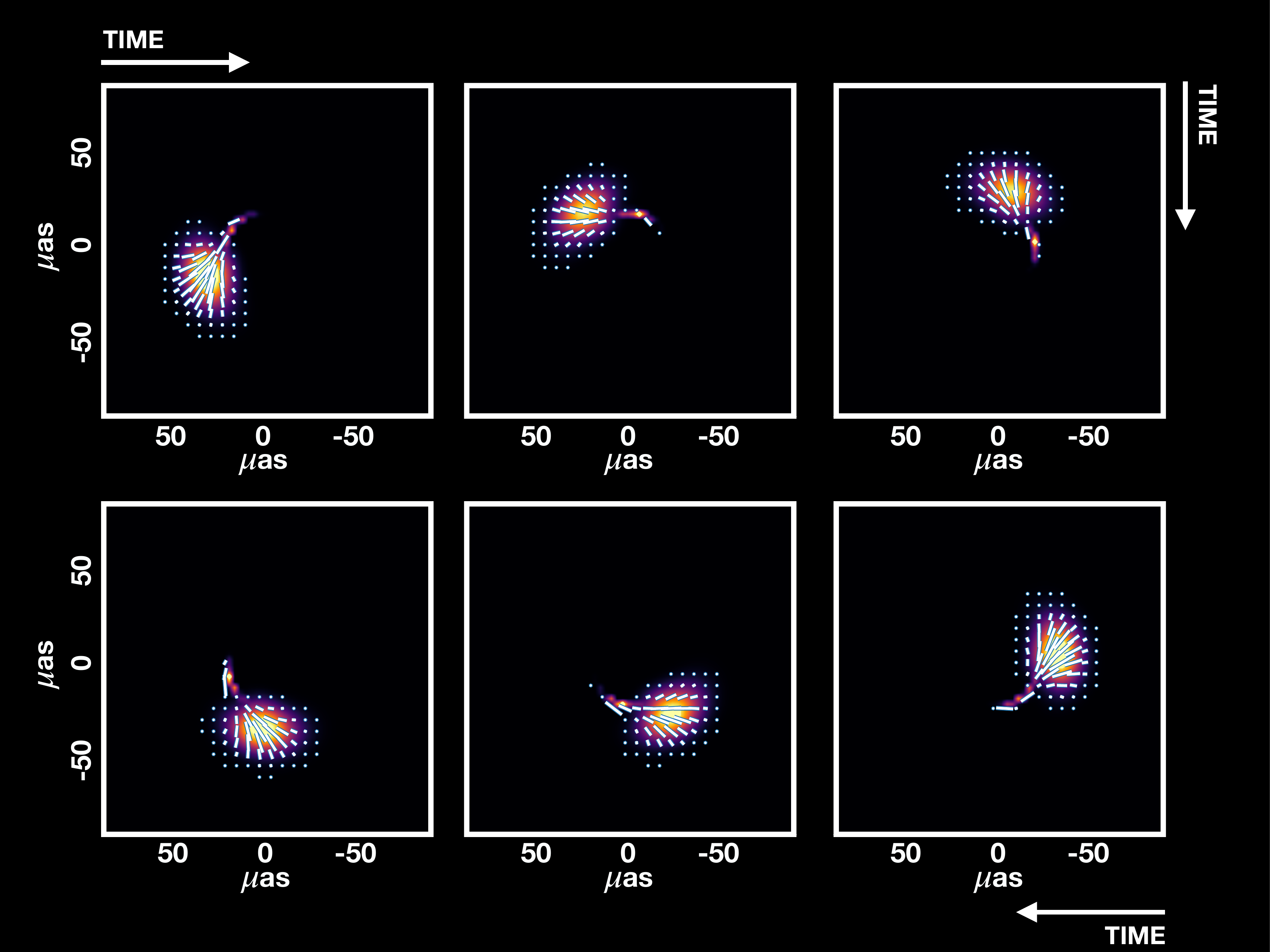}
   \caption{Snapshots of the hotspot as it orbits the black hole clockwise on sky in a vertical magnetic field. The orbital radius is eight gravitational radii. Total intensity is shown as false colour in the background. Polarisation direction is shown as white ticks in the foreground. Their length is proportional to the linear polarisation fraction in that pixel. The hotspot samples the magnetic field geometry in time as it moves along the orbit, so that the time-resolved polarisation encodes information about the spatial structure of the magnetic field.}
   \label{fig:hotspot_snapshots}%
    \end{figure*}

Example snapshots of a hotspot model in a vertical field ($\theta=0$) and the resulting polarisation configuration are shown in Figure  \ref{fig:hotspot_snapshots}. The effects of lensing can be appreciated in the form of secondary images. It can be seen as well that as the hotspot moves along its orbit around the black hole, it samples the magnetic field geometry in time, so that the time-resolved polarisation encodes information about the spatial structure of the magnetic field.

   \begin{figure*}
   \centering
    \includegraphics[trim = 0 0 0cm 0, clip=false,width=0.6\textwidth]{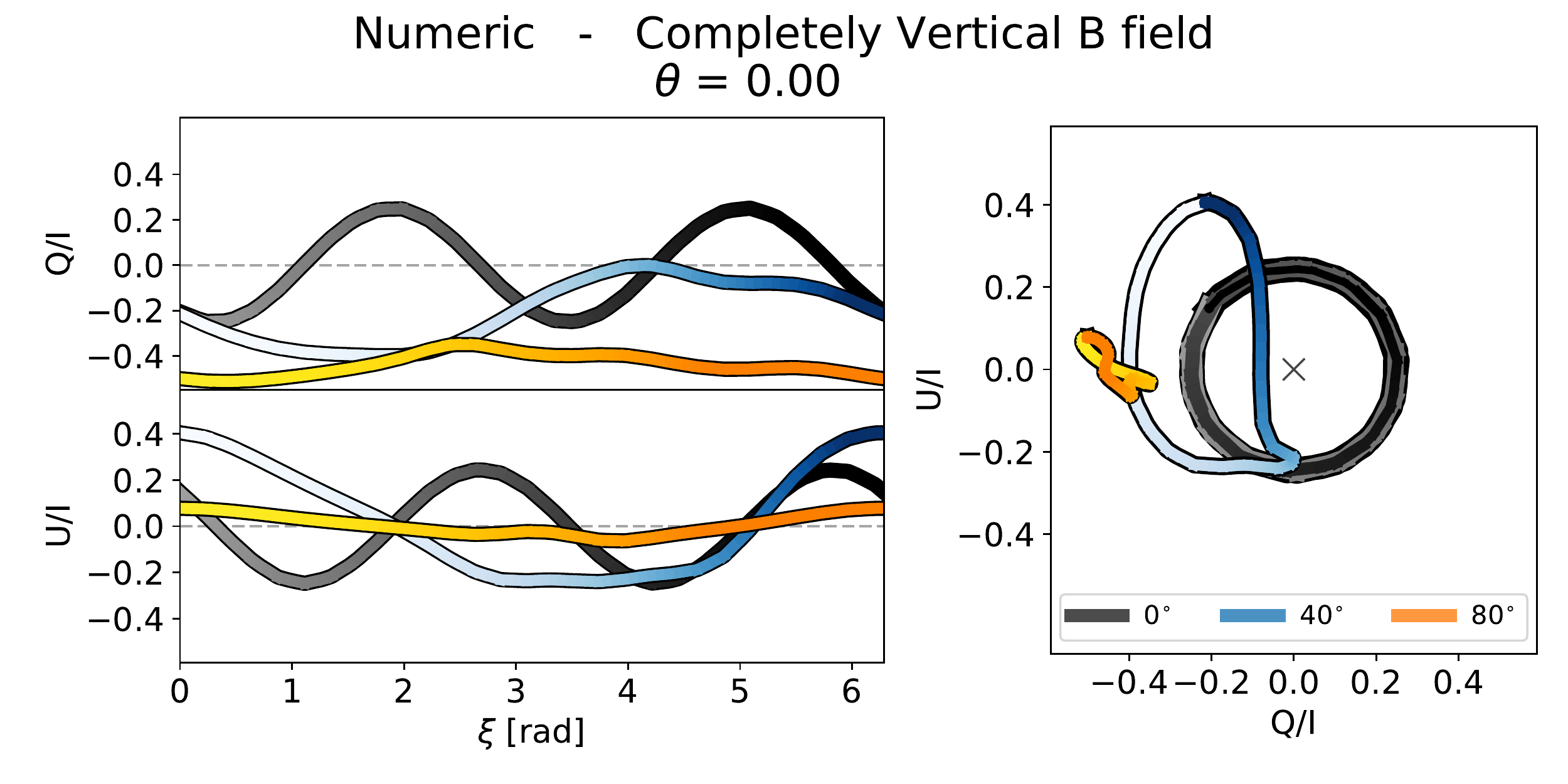}
    \includegraphics[trim = 0 0 0cm 0, clip=false,width=0.6\textwidth]{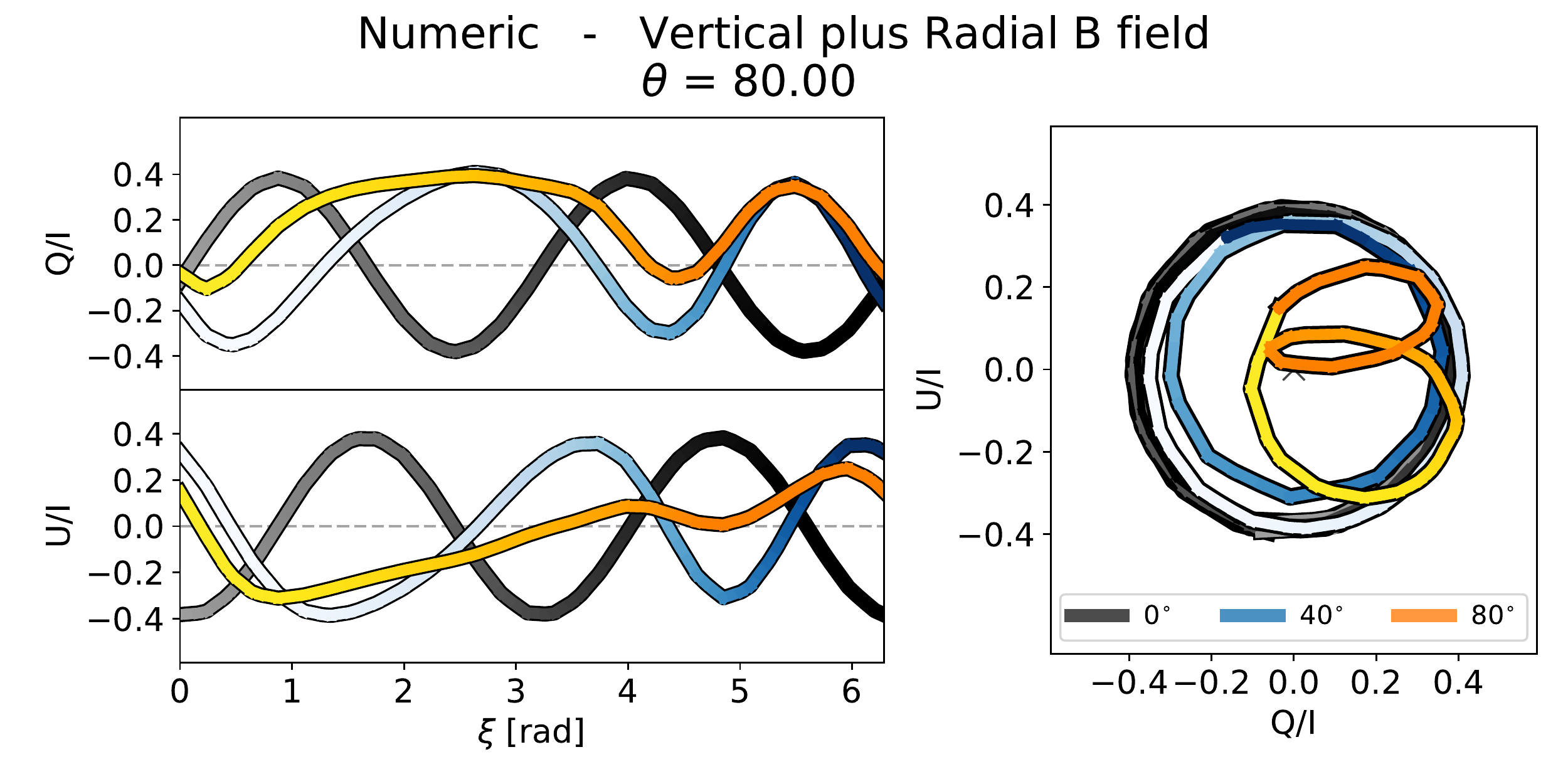}
   \caption{Ray-tracing calculations of the linear Stokes parameters $Q$ and $U$ in a vertical plus radial magnetic field with the same $\theta$ as those in the analytic model. 
   The coarse $QU$ loops are due to the time sampling in our simulations. 
   Top: magnetic field inclination of $\theta=0$ (completely vertical). Bottom: significantly radial magnetic field ($\theta=80$).
   In contrast to the analytic case, numerical calculations in a completely vertical magnetic field at low inclinations show that $Q$ and $U$ oscillate in time and trace loops in $QU$ space due to light bending. }
   \label{fig:numeric_examples}%
    \end{figure*}

Figure  \ref{fig:numeric_examples} shows the numeric calculations of hotspot models with the same magnetic field angles as those in the analytic approximation. 
Inclination and $\theta$ are key parameters in the observed number and shape of $QU$ loops. 
In contrast to the analytic case, in a significantly vertical field ($\theta\simeq0$, top of Fig. \ref{fig:numeric_examples}), the polarisation is not zero. This is mainly due  to light bending, which introduces an effective radial component to the wave-vector in the plane of the observer's camera. This radial component of $\hat{k}$ leads to an additional azimuthal contribution to $\bar{P}$.
The $\theta=0$ cases show that this effect alone is able to generate $QU$ loops. We see again that increasing inclination leads to a change from two $QU$ loops per hotspot revolution at low inclinations to a single $QU$ loop at high inclinations.

The cases where $\theta \xrightarrow{}90^\circ$ (bottom of Fig. \ref{fig:numeric_examples}) show that increasing this parameter also leads to scenarios with two $QU$ loops per hotspot orbit. The shape of the numerical $Q$ and $U$ curves is similar to the analytic versions. The differences are due to the inclusion of relativistic effects in the ray-tracing calculations. We note that numerical models with a vertical plus toroidal magnetic field show similar features and behaviour to those in the vertical plus radial case (see Appendix \ref{appendix:ver+tor_kerr}).

\section{Model fitting}
\label{sec:fitting}
We calculated normalised Stokes parameters $Q/I$ and $U/I$ from ray tracing simulations of a grid of hotspot models, folded them through the instrumental response (Eq. \ref{eq:MFT}), and compared them to GRAVITY's measured $Q'/I'$ and $U'/I'$. The parameters of the numerical model are the orbital radius $R_0$, the size of the hotspot $R_{\rm spot}$, the viewing angle $i$, and the tilt angle of the magnetic field direction $\theta$.

We understand qualitatively how the hotspot size and the orbital radius affect the $Q$ and $U$ curves. 
`Smoother' curves, where the amplitude of the oscillations is reduced, are produced either with increasing hotspot sizes at fixed orbital radius or with decreasing $R_0$ at a fixed hotspot size, due to beam depolarisation (see Appendix \ref{appendix:LP_discussion}).
Since performing full ray tracing simulations is computationally very expensive, and due to the fact that the curves change smoothly and gradually with $R_0$ and $R_{\rm spot}$, we chose to fix their values to $R_0=8 R_g$ and $R_{\rm spot}=3 R_g$, $R_g$ the gravitational radius. We then scaled them in both period and amplitude to match the data better in the following manner.

Given the duration of a flare $\Delta t$, we could scale a hotspot's period by a factor $nT$ to set the fraction of orbital periods that fit into this time window. The new radius of the orbit is then $R\propto (\Delta t/nT)^{2/3}$. This rescaling introduced small changes in fit quality compared to re-calculating new models, within our parameter range of interest (see Appendix \ref{appendix:radii_comparison}).
We absorbed the effect of beam depolarisation into a factor $s$ that scales the overall amplitude of both $Q$ and $U$ and, therefore, the linear polarisation fraction as well. 

Given a hotspot's period, the relative phase reflects the hotspot position relative to an initial position measured at some initial time, where the phase is defined to be zero.
We chose the initial position of the hotspot based on the astrometric measurement of the orbital motion of the flare in \cite{gravity2020_michi}. Specifically, we chose the initial phase $\xi$ to match the initial position of the best-fit orbital model to the astrometry. 

\subsection{Application to the July $28^{\rm{}}$ flare} 

\begin{figure*}
\centering
\includegraphics[trim = 0 0 0cm 0, clip=false,width=1.0\textwidth]{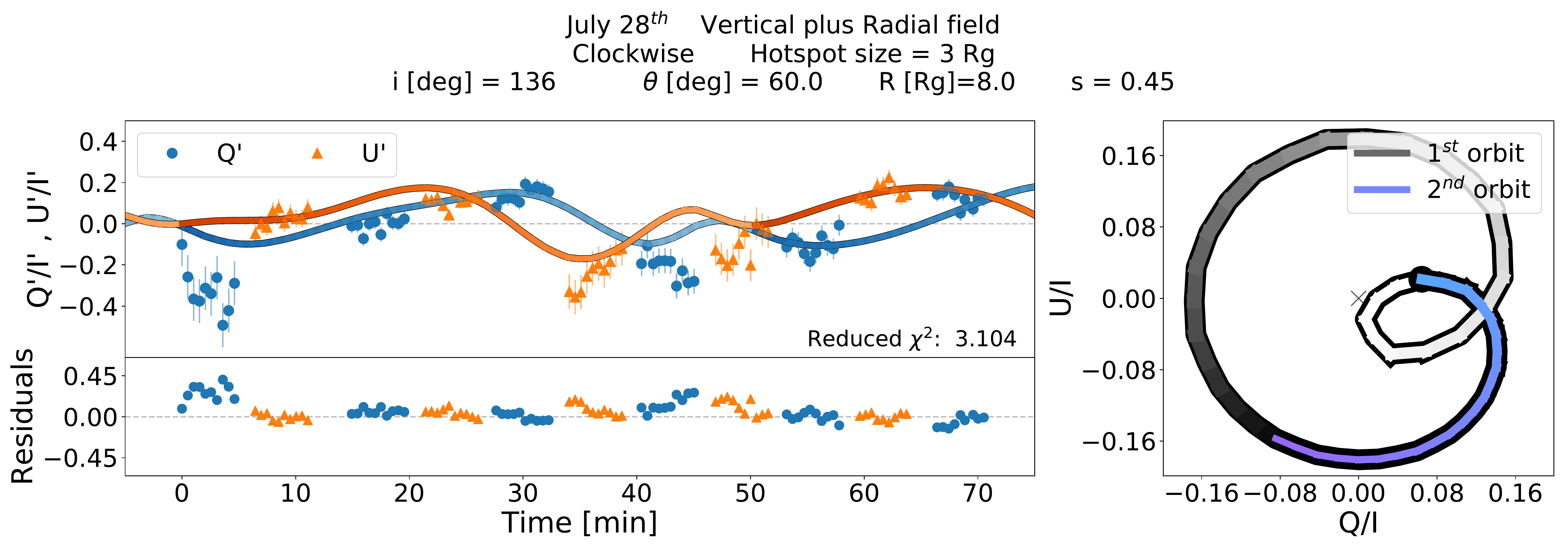}
\caption{Best fit to the July $28^{\rm }$ NIR flare. The colour gradient denotes the periodic evolution of the hotspot along its orbit, moving from darker shades to lighter as the hotpot completes one revolution. The curves qualitatively reproduce the data. The preferred parameter combination favours a radius of $8 \ R_g$ and both moderate $i$ and $\theta$ values. 
 }
\label{fig:july28_bestfit}
\end{figure*}

The observed $Q'/I'$ and $U'/I'$ were measured from fitting interferometric binary models to GRAVITY data. The binary model measures the separation of Sgr A* and the star S2, which were both in the GRAVITY interferometric field of view ($\simeq 50$ mas) during 2018. For more details, see \citet{gravity2020_sebo}. We measured polarisation fractions assuming that S2's NIR emission is unpolarised. The 70 minute time period analysed is limited by signal-to-noise: binary signatures are largest when Sgr A* is brightest. As a result, we focused on data taken during the flare. We fitted to data binned by $30$ seconds since the flux ratio can be rapidly variable. We further adopted error bars on polarisation fractions using the rms of measurements within $300$s time intervals since direct binary model fits generally have $\chi^2 > 1$, and as a result underestimate the fit uncertainties.

We computed a grid of models with $i$, $\theta$, $s$, and $nT$ as parameters: $i \in [0-180]$ in increments of $\Delta i=4^\circ$; $\theta \in [0-90)$, $\Delta \theta=4^\circ$; $s \in [0.4-0.8]$, $\Delta s=0.05$, and $nT$ such that the allowed range of radii for the fit is $R=8-11\ R_g$ with $\Delta R=0.2$. 
We have included this prior in radii to match the constraint from the combined astrometry of the three bright GRAVITY 2018 flares \citep{gravity2020_michi}. The best fit parameters and corresponding polarised curves are shown in Figure  \ref{fig:july28_bestfit}. 
We find that the curves qualitatively reproduce the data and that the statistically preferred parameter combination for July $28^{\rm{}}$, with a reduced $\chi^2\sim3.1$, favours a radius of $8\ R_g$ and moderate $i$ and $\theta$ values (left panel of Figure  \ref{fig:july28_bestfit}). 
In $QU$ space, these parameters produce two intertwined and embedded $QU$ loops of very different amplitudes in time (right panel of Figure  \ref{fig:july28_bestfit}). The outer one is fairly circular, centred approximately around zero and with an average radius of $0.18$. The inner one has a horizontal oblate shape 
with a $QU$ axis ratio of approximately 2:1, does not go around zero, and represents a much smaller fraction of the orbit than the larger loop.
These moderate values of $\theta$ imply that a magnetic field with significant components in both the radial and vertical directions is favoured. 

The hotspot is free to trace a clockwise ($i>90^\circ$) or counterclockwise ($i<90^\circ$) motion on-sky. At fixed $\theta$, this change in apparent motion results in an inversion of the order in which the maxima of the $Q$ and $U$ curves appear\footnote{ This is also equivalent to an inversion of the curve in time and does not modify the features of the curve.}. 

This effect is due to relativistic motion \citep{blandford1979,bjornsson1982}. When the magnetic field is purely toroidal (velocity parallel to $\bar{B}$), the polarisation angle is independent of velocity. When there is a field component perpendicular to the velocity (poloidal field), relativistic motion induces an additional swing of the polarisation angle in the direction of movement where magnitude depends on the velocity. We ignore this effect in the analytic approximation above, but it is included in our numerical calculations. 

The data favour models where the maxima in $U'/I'$ precede those of $Q'/I'$. This behaviour is observed in the case of clockwise motion ($i>90^\circ$) with $\theta \in [0^\circ-90^\circ]$ and in counterclockwise motion ($i<90^\circ$) with $\theta \in [90^\circ-180^\circ]$. In fact, model curves at a given $i>90^\circ$ and $\theta \in [0^\circ-90^\circ]$ are identical to those with their `mirrored' values $i'=180^\circ-i$ and $\theta'=180^\circ-\theta$. In our analysis, we consider $\theta \in [0^\circ-90^\circ]$, which favours a clockwise motion. However, we cannot uniquely determine the apparent direction of motion of the hotspot due to this degeneracy. 

Our models overproduce the observed linear polarisation fraction by a factor of $\sim1.7$ (scaling factor $s\simeq0.4<1$). The maximum observed polarisation fraction is $\simeq 30\%$, while it is $\simeq 50\%$ in our models. The degree of depolarisation introduced by the VLTI is not substantial enough to reduce the model linear polarisation fraction to the observed one. Moreover, in the NIR, there are no significant depolarisation contributions from absorption or Faraday effects. As a result, we conclude that the low observed polarisation fraction is likely the result of beam depolarisation. The observed low polarisation fraction implies that the flare emission region is big enough to resolve the underlying magnetic field structure. In the context of our model, this could imply a larger spot size. It could also indicate a degree of disorder in the background magnetic field structure, for example as a  result of turbulence.

\subsection{Application to the July $22^{\rm{}}$ flare}

\begin{figure*}
\centering
\includegraphics[trim = 0 0cm 0cm 0cm, clip=false,width=1.0\textwidth]{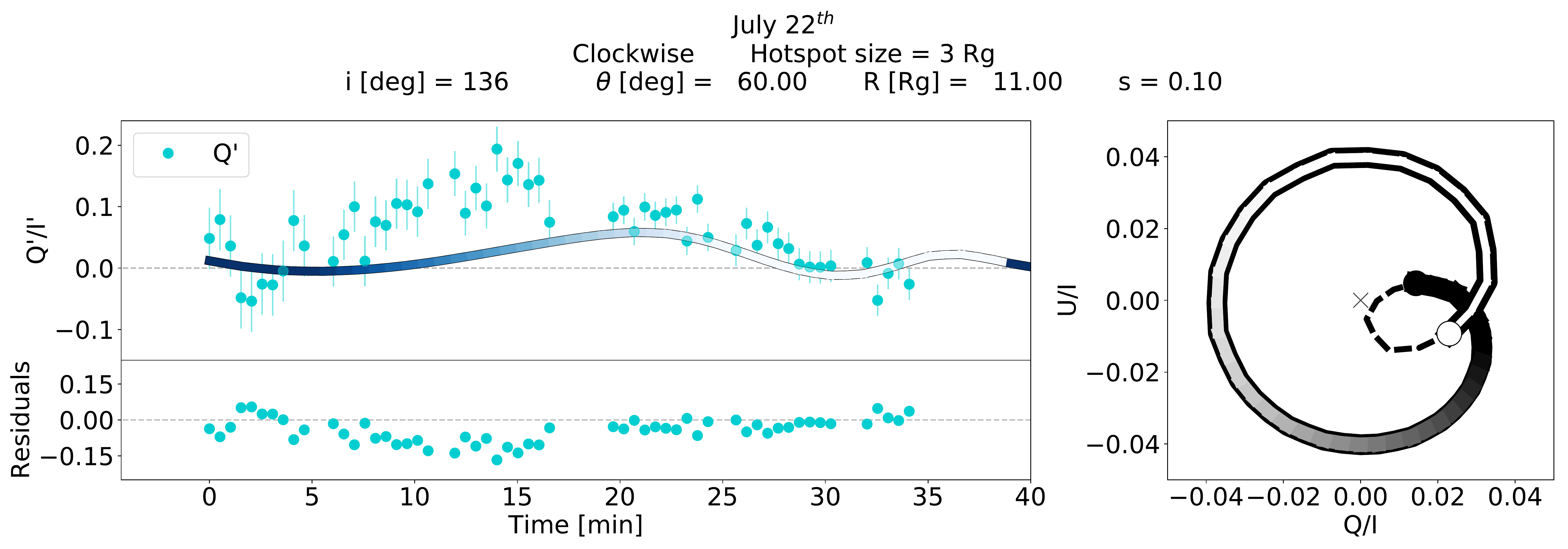}
\caption{Fit to the July $22^{\rm }$ NIR flare without restricting the phase difference between this night and that of July $28^{\rm }$. The colour gradient denotes the evolution of the hotspot as it completes one revolution. The viewer's inclination, magnetic field geometry, and orbital direction have been fixed to the values found for the July $28^{\rm }$ flare. The fit favours values of $R_0\sim 11\ R_g$ and there is no initial phase difference between the nights (no difference in starting position on-sky), which is out of the allowed uncertainty range for the astrometry. 
  }
\label{fig:july22_fit}
\end{figure*}

July $28^{\rm }$ is the only night with an observed infrared flare in which GRAVITY recorded both Stokes $Q'$ and $U'$. Since a single polarisation channel is insufficient to constrain the full parameter space used in our numerical models, we restricted ourselves to the night of July $22^{\rm }$, as this observation has the highest precision astrometry\footnote{The astrometry of the May $27^{\rm }$ and June $27^{\rm }$ flares is not good enough to pin point their starting location on sky, so it is not at all possible to restrict the phase difference between them and the July $28^{\rm }$ flare.}, and fixed the viewer inclination and magnetic field geometry to be the same as the best fit model to the July $28^{\rm }$ data. We scaled the curves in amplitude with $s \in [0.05-0.35]$, $\Delta s=0.05$.

The initial position on sky for both flares is constrained by astrometric data and, therefore, so is the phase offset between both curves. 
With a fixed phase difference between the curves and free range of radii, we find that the July $22^{\rm }$ data favours extremely large values of $R_0>20 \ R_g$, which are outside of the allowed range obtained from astrometric measurements.
In allowing the phase difference to be free and constraining the radii to $8-11 \ R_g$, with $\Delta R=0.2$, we find that the data tend to values of $R_0\sim11 \ R_g$ and a phase difference between curves of $0^\circ$ (Figure  \ref{fig:july22_fit}).
This phase difference value (and position difference associated with it) is outside of the allowed uncertainties in the initial position indicated by the astrometric data. 
The fact that the magnetic field parameters that describe the July $28^{\rm{}}$ flare fail to adequately fit the data from July $22^{\rm{}}$ may indicate that the background magnetic field geometry changes on a several-day timescale.

\section{Summary and discussion}
\label{sec:discussion}

In this work, we present an extension of the initial analysis of polarisation data performed in \citet{gravity2018flare}. We forward modelled $Q$ and $U$ Stokes parameters obtained from ray-tracing calculations of a variety of hotspot models in different magnetic field geometries, transformed them into quantities as seen by the instrument, and fitted them directly to the polarised data taken with GRAVITY.

This allowed us to not only fit data directly without making assumptions about Stokes $V$ or the interpolation of data in non-simultaneous $Q$ and $U$ measurements, but also to predict the behaviour in time of the polarised curves and loops for the cases where only one of the parameters was measured.

We have shown that the hotspot model serves to qualitatively reproduce the features seen in the polarisation data measured with GRAVITY.
A moderate inclination and moderate mix of both vertical and radial fields provide the best statistical fit to the data.
Consistent results are found by fitting the data with a vertical plus toroidal field component (Appendix \ref{appendix:ver+tor_kerr}). We note that this result does not rely on the assigned strength of the magnetic field, since the model curves are scaled in amplitude, but rather it is only from the geometry of the field. Magnetic fields with a non-zero vertical component fit the data statistically better. This supports the idea that there is some amount of ordered magnetic field in the region near the event horizon with a significant poloidal field component. The presence of this component is associated with magnetic fields that are dynamically important and it confirms the previous finding of strong fields in \citet{gravity2018flare}. Spatially resolved observations at $1.3$mm also found linear polarisation structure consistent with a mix of ordered and disordered magnetic field \citep{johnson2015}.

Matching the clockwise direction of motion inferred by the astrometric data would require that $\theta \in [0^\circ-90^\circ]$. 
Under this assumption, the results are also in accordance with the angular momentum direction and orientation of the clockwise stellar disc and gas cloud G2 \citep{bartko2009,gillessen2019,pfuhl2015,plewa2017}.

We have chosen the bright NIR flare on July $28^{\rm }$, 2018 since it is the only one for which both linear Stokes parameters have been measured. Naturally, increasing the number of full data sets in future flares will be useful in constraining the parameter range more. 

Our models overproduce the observed NIR linear polarisation fraction of $\sim30\%$ by a factor of $\sim 1.7$, and they must be scaled down to fit the data.
In the compact hotspot model context, this implies that an emission region size larger than $3 \ R_g$ is needed to depolarize the NIR emission through beam depolarisation. Including shear in the models would naturally introduce depolarisation since a larger spread of polarisation vector directions (or equivalently, the magnetic field structure) would be sampled at any moment \citep[e.g.][]{gravity2020_michi,tiede2020}. However, this might smooth out the fitted curves and would probably change the fits. 
In any case, the observed low NIR polarisation fraction means that the observed emission region resolves the magnetic field structure around the black hole.

Though simplistic, the hotspot model appears to be viable for explaining the general behaviour of the data.  
It would be interesting to study the polarisation features of more complex, total emission scenarios explored in other works. \citet{ball2020} study orbiting plasmoids that result from magnetic reconnection events close to the black hole, where some variability in the polarisation should be caused by the reconnecting field itself. \citet{dexter2020_MAD} find that material ejected due to the build-up of strong magnetic fields close to the event horizon can produce flaring events where the emission region follows a spiral trajectory around the black hole. In their calculations, ordered magnetic fields result in a similar polarisation angle evolution as we have studied here. Disorder caused by turbulence reduces the linear polarisation fraction to be consistent with what is observed.

Spatially resolved polarisation data are broadly consistent with the predicted evolution in a hotspot model. This first effort comparing these types of models directly to GRAVITY data shows the promise of using the observations to study magnetic field structure and strength on event horizon scales around black holes.

\begin{acknowledgements}
JD is pleased to thank D.P. Marrone, J. Moran, M.D. Johnson, G.C. Bower, and A.E. Broderick for helpful discussions related to signatures of orbital motion around black holes from polarized synchrotron radiation. We thank the anonymous referee for their constructive comments. This work was supported by a CONACyT/DAAD grant (57265507) and by a Sofja Kovalevskaja award from the Alexander von Humboldt foundation. A.A. and P.G. were supported by Funda\c{c}\~{a}o para a Ci\^{e}ncia e a Tecnologia, with grants reference UIDB/00099/2020 and SFRH/BSAB/142940/2018.

\end{acknowledgements}

\bibliographystyle{aa}
\bibliography{GRAVITY_flare_pol}

\begin{appendix} 

\section{Vertical plus radial field in Boyer-Lindquist coordinates}
\label{appendix:ver+rad_kerr}

In the Boyer-Lindquist coordinate frame, a magnetic field with a vertical plus radial components can be written as:

\begin{align}
\underline{B}&=(B^t,\ B^r,\ B^\theta,\ B^\phi) \nonumber \\
             &= (B^t,\ \delta_c B^\theta,\ B^\theta,\ 0),
\label{eq:b_gr}
\end{align}

\noindent where $B^\mu$ are the contravariant components of $\underline{B}$ and $\delta_c\equiv B^r/B^\theta$. 
The magnetic field must satisfy the following conditions:
\begin{align}
    B_\mu u^\mu &= g_{\mu\nu}B^\nu u^\mu = 0 \nonumber\\
    B_\mu B^\mu &= g_{\mu\nu}B^\nu B^\mu = B^2,
    \label{eq:B_norm}
\end{align}

\noindent where $u^\mu$ are the contravariant components of the four-velocity, $B$ is the magnitude of $\underline{B}$, and $g_{\mu\nu}$ are the covariant components of the Kerr metric. 
In Boyer-Lindquist coordinates with $G=c=M=1$, the non-zero components of the metric are:

\begin{align}
g_{tt} &=-\left(1-\frac{2r}{\Sigma}\right) \nonumber  \\ 
g_{rr} &= \frac{\Sigma}{\Delta}   \nonumber \\    
g_{\theta\theta}&=\Sigma  \nonumber \\
g_{t\phi}&=g_{\phi t}=-\frac{2r}{\Sigma}a\sin{}^2\theta    \nonumber \\  
g_{\phi\phi} &=\left[r^2+a^2+\frac{2ra^2}{\Sigma}\sin{}^2\theta\right]\sin{}^2\theta,    
\label{eq:kerr_met}
\end{align}
\noindent where
\begin{align}
\Delta &\equiv r^2-2r+a^2 \nonumber \\
\Sigma &\equiv r^2+a^2\cos{}^2\theta,  \nonumber 
\end{align}

\noindent where $a$ is the dimensionless angular momentum of the black hole.

Using Eq. (\ref{eq:b_gr}), (\ref{eq:B_norm}), and (\ref{eq:kerr_met}), it follows that the Boyer-Lindquist coordinate frame contravariant components of the magnetic field are

\begin{align}
    B^{t}&= - C B^\theta ; \nonumber \\
    B^{r}&=  \delta_c B_\theta = B_\theta \ \delta_{LNRF} /r ;\nonumber \\
    B^{\theta}&= B \  (g_{tt}C^2+g_{rr}\delta_c^2 +g_{\theta\theta})^{-1/2} ; \nonumber \\
    B^{\phi}&= 0 ; \nonumber\\
    \rm{with} \nonumber \\
    C &\equiv \frac{\delta_c g_{rr}u^r+g_{\theta\theta}u^\theta}{g_{tt}u^t+g_{t\phi}u^\phi} () \label{eq:B_LNRF} \\
    \rm{and} \nonumber\\
    \delta_c &= \delta_{LNRF}/r \ \ \ ; \ \ \  \delta_{LNRF}=\frac{B^{(r)}}{B^{(\theta)}} \nonumber
\end{align}
\noindent where $\delta^{LNRF}$ is the ratio of the radial and poloidal magnetic field components in the locally non-rotating frame \citep[LNRF, ][]{bardeen1973} and $B^{(\mu)}$ are the contravariant components of $\underline{B}$ in the LNRF:

\begin{align}
    B^{(t)}&= (\Sigma \Delta/ A)^{1/2} B^t \sim B^t; \nonumber \\
    B^{(r)}&=  (\Sigma/\Delta)^{1/2} B^r \sim B^r; \nonumber \\
    B^{(\theta)}&= \Sigma^{1/2} B^\theta \sim rB^\theta ;  \nonumber \\
    B^{(\phi)}&= -\frac{2ra\sin{\theta}}{(\Sigma A)^{1/2}}B^t+(A/\Sigma)^{1/2}\sin{\theta}B^\phi \sim r\sin\theta B^\phi; \label{eq:LNRF}
    \\
    \rm{with} \nonumber \\
    A &\equiv (r^2+a^2)^2-a^2\Delta\sin{}^2\theta, \nonumber
\end{align}
\noindent where the expression to the far right is obtained by assuming $r \gg a$ (as it is in the hotspot case). The variable $\delta$ used in the main text (Eq. (\ref{eq:b_ver_analytic})) corresponds to $\delta_{\rm LNRF}$ defined here as being calculated using the $r \gg a$ approximation.

\section{Analytic approximation with a vertical plus toroidal magnetic field}
\label{appendix:toroidal_B}

\begin{figure*}
\includegraphics[trim = 0 0 0cm 0, clip=false,width=0.49\textwidth]{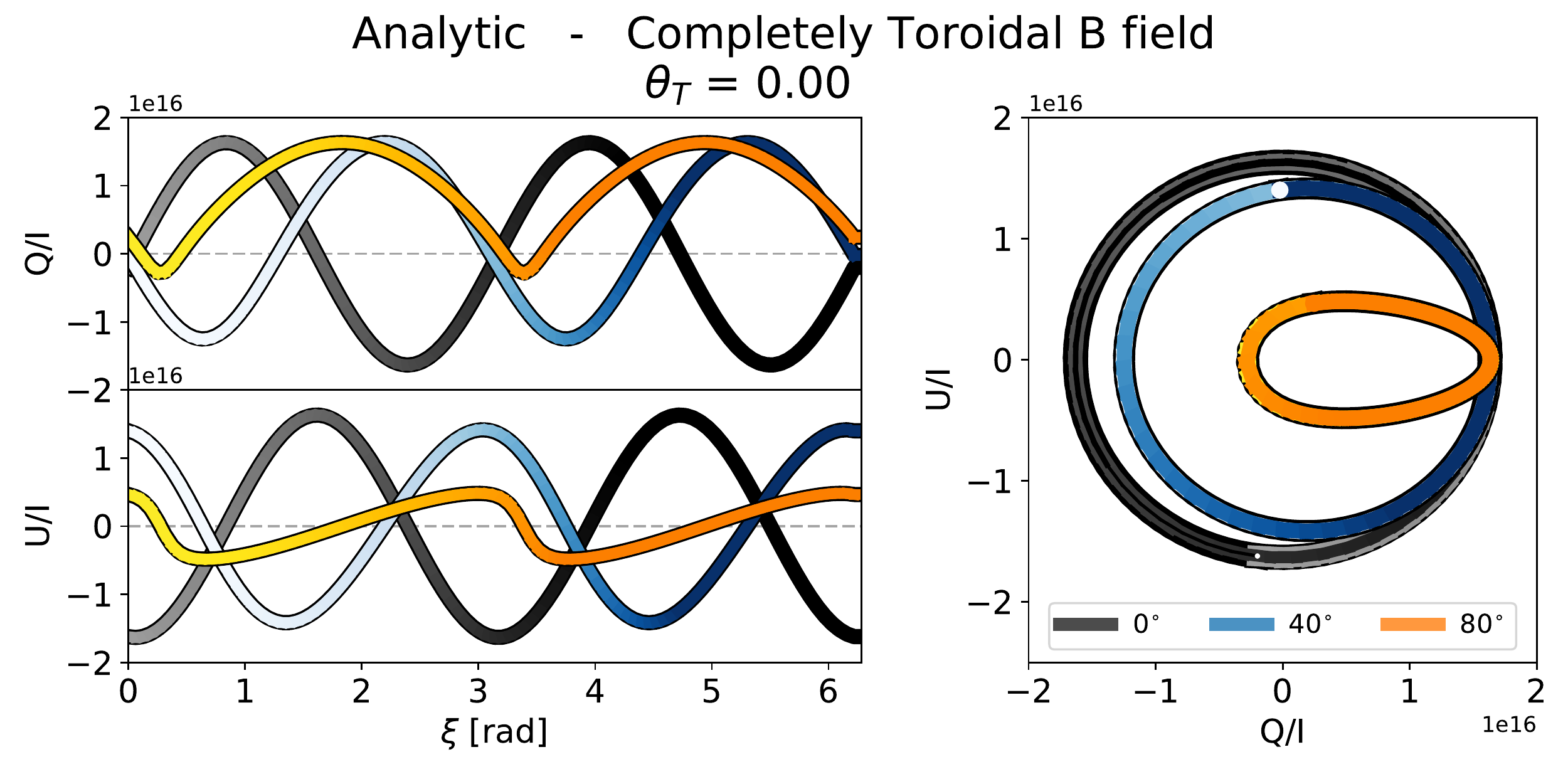}
\includegraphics[trim = 0 0 0cm 0, clip=false,width=0.49\textwidth]{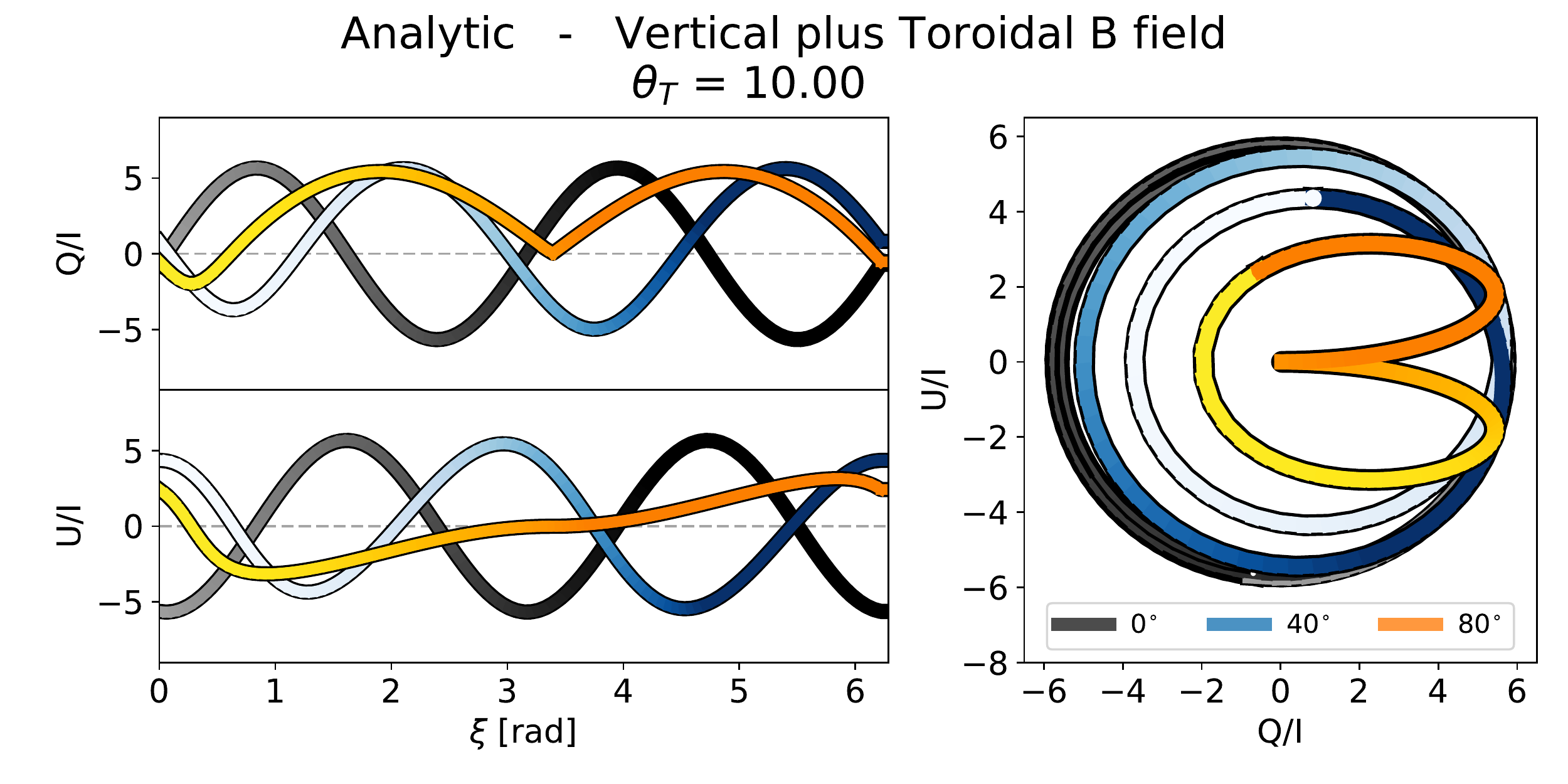}
\includegraphics[trim = 0 0 0cm 0, clip=false,width=0.49\textwidth]{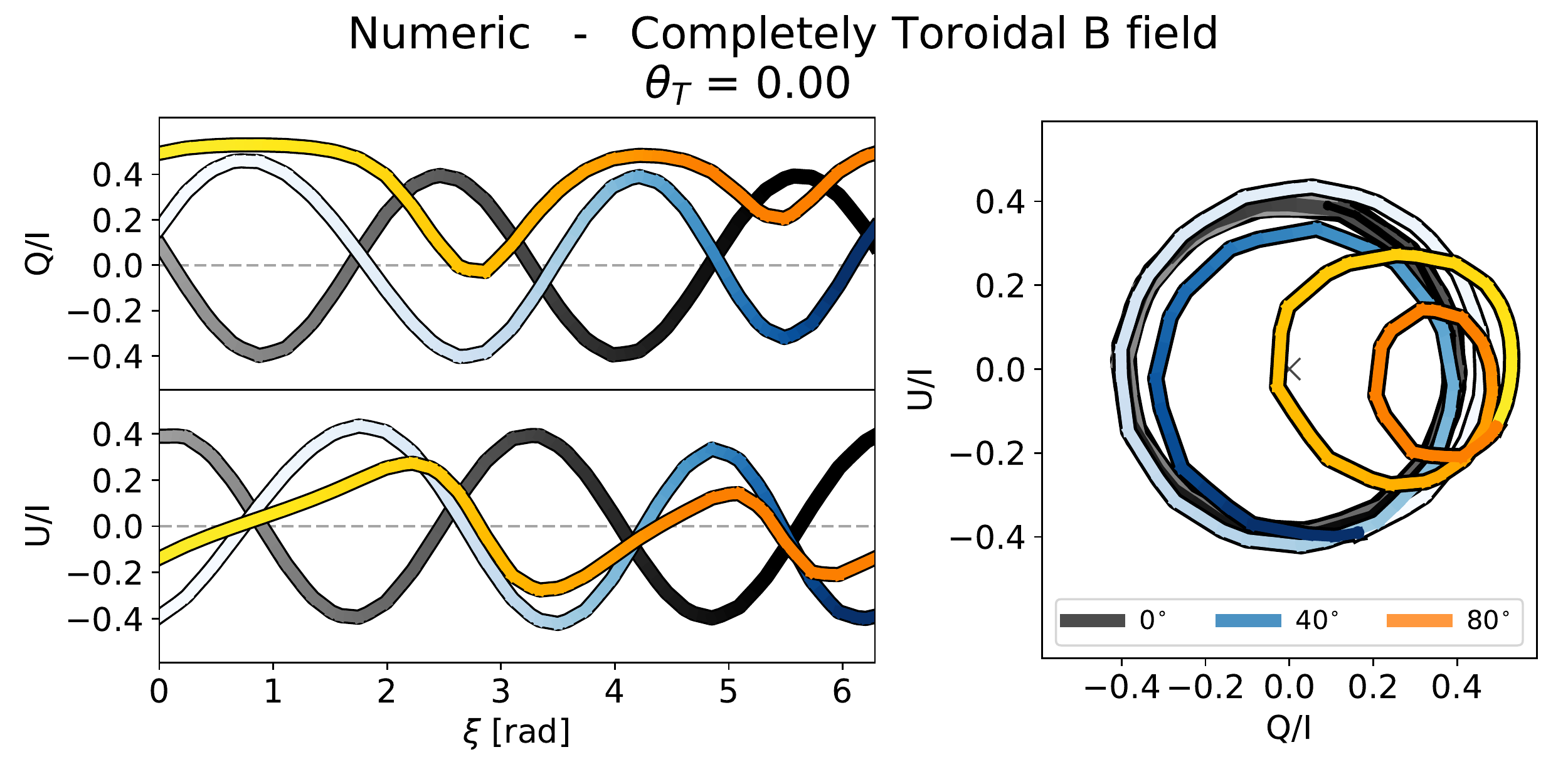}
\includegraphics[trim = 0 0 0cm 0, clip=false,width=0.49\textwidth]{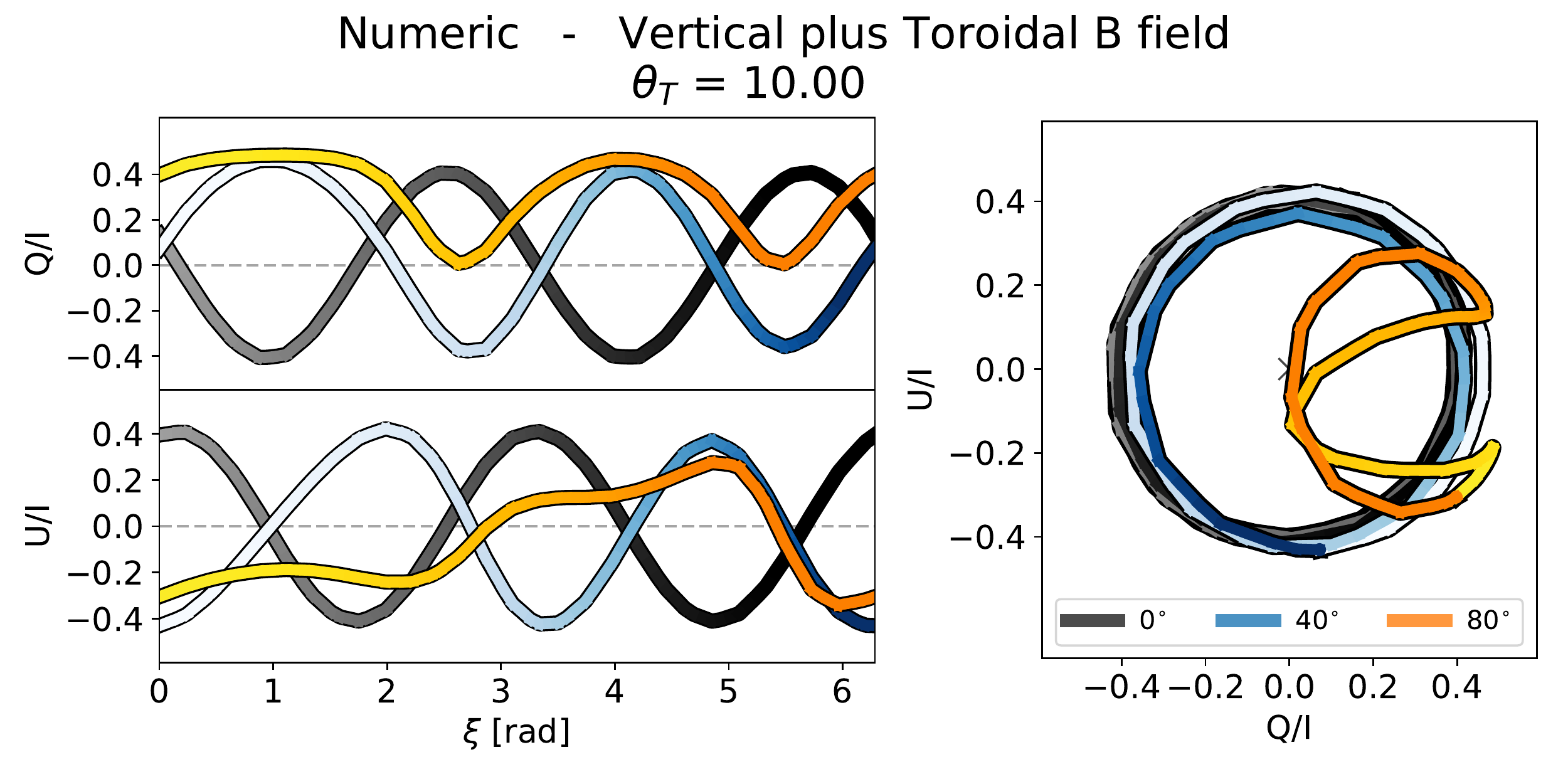}
\caption{Analytic and ray-tracing calculations of $Q$ and $U$ curves in the case of a toroidal magnetic field. Two loops are always observed. In the case of the analytic case (top), both are superimposed. This is broken by the accounting for light bending in the ray-tracing calculations (bottom). It can also be seen that toroidal and completely radial configurations produce the same curves, save for a a scaling factor and a phase offset.}
\label{fig:toroidal_numeric}
\end{figure*}

In the case of a vertical plus toroidal magnetic field, the magnetic field can be written as $\bar{B} \propto \hat{z}+\lambda \hat{\phi}$, where $\lambda\propto \tan{\theta_T}$ is the strength of the toroidal component, $\theta_T$ is the angle measured from the toroidal component to the vertical component ($\theta_T=0$ denotes a completely toroidal field), and

\begin{equation}
 \hat{\phi}=  -\sin{\xi}\ \hat{\alpha}+\cos{i}\cos{\xi}\ \hat{\beta}  -\sin{i}\cos{\xi}\ \hat{k}
 \label{eq:phi_vector}
 \end{equation}
 
\noindent is the canonical vector in the azimuthal direction (Figure  \ref{fig:hotspot_diagram}). 
We note that $\hat{r}\cdot\hat{\phi}=0$. 

The polarisation vector in a flat space given by $\hat{k}\times\bar{B}$ is then

\begin{equation}
\bar{P} \propto -(\sin{i}+\lambda\cos{i}\cos{\xi})\ \hat{\alpha}-\lambda\sin{\xi}\ \hat{\beta}
\label{eq:P_toroidal}
\end{equation}

\noindent and the polarisation angle is given by
\begin{equation}
\psi = \tan^{-1} \left( \frac{ \lambda \sin{\xi}}{\sin{i}+\lambda\cos{i}\cos{\xi}} \right).
\label{eq:pol_angle_tor}
\end{equation}

It can be seen from expression (\ref{eq:P_toroidal}) that at low inclinations or when $\lambda>>1$ (complete toroidal magnetic field), the polarisation has a radial configuration ($\bar{P}\propto\hat{r}$, Eq. \ref{eq:r_vector}). This is geometrically equivalent to the polarisation having a toroidal configuration (similar to the one generated by a completely radial magnetic field, see Section \ref{sec:model}) 
with a phase offset of $\pi/2$ in $Q$ and $U$. 
In this case, we would expect to have two superimposed $QU$ loops in one revolution of the hotspot.

Figure \ref{fig:toroidal_numeric} shows a comparison between the analytic (top) and numeric (bottom) calculations for a vertical plus toroidal magnetic field (Appendix \ref{appendix:ver+tor_kerr}). As expected, in the analytic case,  there are always two superimposed loops in $QU$ space in the case of a completely toroidal field.
In the numeric calculations, this is also the case given that light bending favours the presence of loops. 
As a vertical component in the field is introduced, the loops no longer overlay on each other. This effect increases with viewer inclination.  
It can also be seen that the completely toroidal and radial cases produce the same $Q$ and $U$ curves at low inclinations, save for a phase offset and scaling factor.

\section{Vertical plus toroidal field in Boyer-Lindquist coordinates}
\label{appendix:ver+tor_kerr}

In the Boyer-Lindquist coordinate frame, a magnetic field with a vertical plus toroidal components can be written as:

\begin{align}
\underline{B}&=(B^t,\ B^r,\ B^\theta,\ B^\phi) \nonumber \\
             &= (B^t,0,\ \eta_c B^\theta,B^\phi)
\label{eq:b_tor}
\end{align}

\noindent where $B^\mu$ are the contravariant components of $\underline{B}$ and $\eta_c\equiv B^\theta/B^\phi$. 
Just as in the vertical plus radial case, the magnetic field must satisfy Eqs. (\ref{eq:B_norm}).

Using Eqs. (\ref{eq:b_tor}), (\ref{eq:B_norm}), and (\ref{eq:kerr_met}), it follows that the Boyer-Lindquist coordinate frame contravariant components of the magnetic field are

\begin{align}
    B^{t}&=  B^\theta/C ; \nonumber \\
    B^{r}&=  0 ;\nonumber \\
    B^{\theta}&= \frac{\sqrt{A}}{\rho} \eta_{LNRF} \sin{\theta} (C-\omega) B^{t}; \nonumber \\
    B^{\phi}&= \frac{C\ B} {\sqrt{g_{tt}+2g_{t\phi}C+g_{\theta\theta}\frac{B^\theta}{B^t}+g_{\phi\phi}C^2}} ; \nonumber\\
    \rm{with} \label{eq:Btor_LNRF} \\
    C &\equiv -\frac{g_{tt}u^t+g_{t\phi}u^\phi}{g_{t\phi}u^t+g_{\phi\phi}u^\phi} \ \ \ ; \ \ \  
    \omega = \frac{2ra}{A}; \nonumber \\
\end{align}
\noindent $\eta_{LNRF}= B^{(\theta)}/B^{(\phi)}=\tan{\theta _T}$ the ratio of the poloidal and toroidal magnetic field components in the LNRF (Eq. (\ref{eq:LNRF})), and $\theta_T$ is the angle measured from the toroidal component to the vertical ($\theta_T=0$ implies a completely toroidal field, Appendix \ref{appendix:toroidal_B}).

\begin{figure*}
\centering
\includegraphics[trim = 0 0cm 0cm 0, clip=true,width=0.8\textwidth]{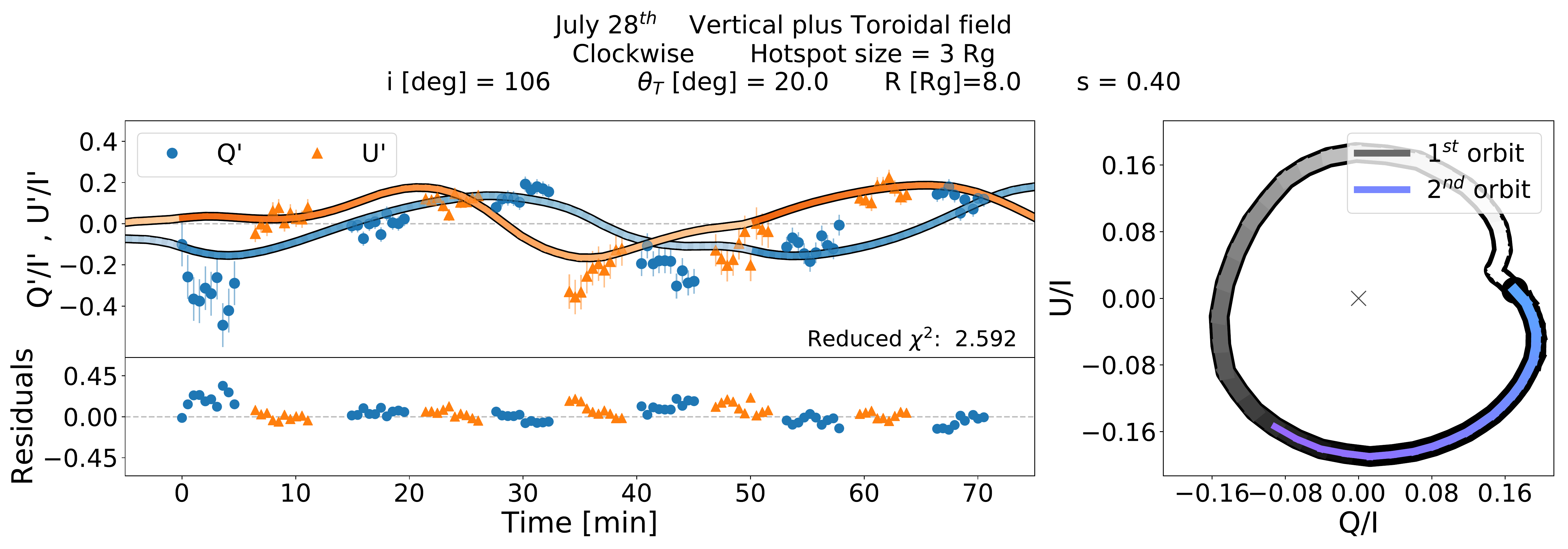}
\caption{Best fit to the July $28^{\rm }$ flare with a vertical plus toroidal magnetic field. The colour gradient denotes the periodic evolution of the hotspot along its orbit, moving from darker shades to lighter as the hotpot completes a revolution. Considering $\theta_T \in [0^\circ-90^\circ]$, a clockwise motion is preferred. The fit has a smaller reduced $\chi^2$ at a slightly higher inclination than the best fit with a vertical plus radial field. The presence of a vertical component in the magnetic field is still required to fit the data better.  }
\label{fig:ver+tor_fit}
\end{figure*}

\begin{figure*}
\centering
\includegraphics[trim = 0 0cm 0cm 0, clip=true,width=0.8\textwidth]{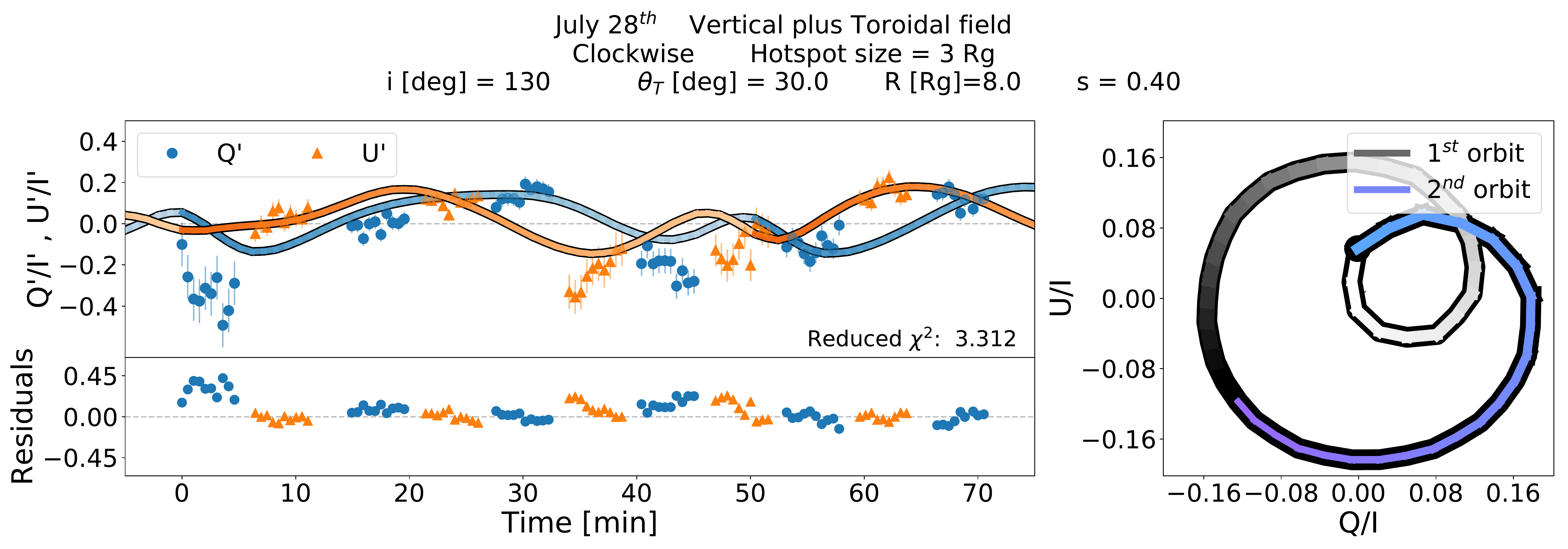}
\caption{Vertical plus toroidal model fit with similar parameters to those of the best fit with a vertical plus radial field. 
 }
\label{fig:ver+tor_fit2}
\end{figure*}

We  fitted the July $28^{\rm }$ data considering this magnetic geometry. Just as in the vertical plus radial case, we computed a grid of models with $i$, $\theta$, $s$, and $nT$ as parameters: $i \in [0-180]$ in increments of $\Delta i=4^\circ$; $\theta_T \in [0-90]$, $\Delta \theta_T=5^\circ$; $s \in [0.4-0.8]$, $\Delta s=0.05$, and $nT$ such that the allowed range of radii for the fit is $R=8-11\ R_g$ with $\Delta R=0.2$. 
The best fit is shown in Figure \ref{fig:ver+tor_fit}. Though a better reduced $\chi^2$ is found at a somewhat higher inclination than the best fit with a vertical plus radial magnetic field (Fig. \ref{fig:july28_bestfit}), the presence of a poloidal component in the magnetic field is still needed. Considering $\theta_T \in [0^\circ-90^\circ]$, a clockwise motion is preferred ($i>90^\circ$). Identical curves can be obtained when the direction of motion is counterclockwise ($i<90^\circ$) and the magnetic field angle is $\theta_T'=180^\circ-\theta_T$. Figure \ref{fig:ver+tor_fit2} presents a model of a vertical plus toroidal magnetic field with similar parameters to those of the vertical plus radial field best fit. 

\section{Spin effects}
\label{apendix:spin_comparison}

\begin{table}
\caption{Reduced $\chi^2$ of best fit of the July $28^{\rm }$ flare data with three dimensionless spins: $0.0,0.9,-0.9$.}
\label{table:spin_comparison}      
\centering                          
\begin{tabular}{c c c}        
\hline
 R  $[\rm{R_g}]$ & $a$ & $\chi^2$\\    
\hline                        
   8.0 & 0.0 & $3.104$ \\      
   8.0 & 0.9 & $3.194$ \\      
   8.0 & $-0.9$ & $3.080$ \\
   \hline
   \hline                                
\end{tabular}
\end{table}

\begin{figure*}
\centering
\includegraphics[trim = 0cm 0 0cm 2cm, clip=true,width=0.8\textwidth]{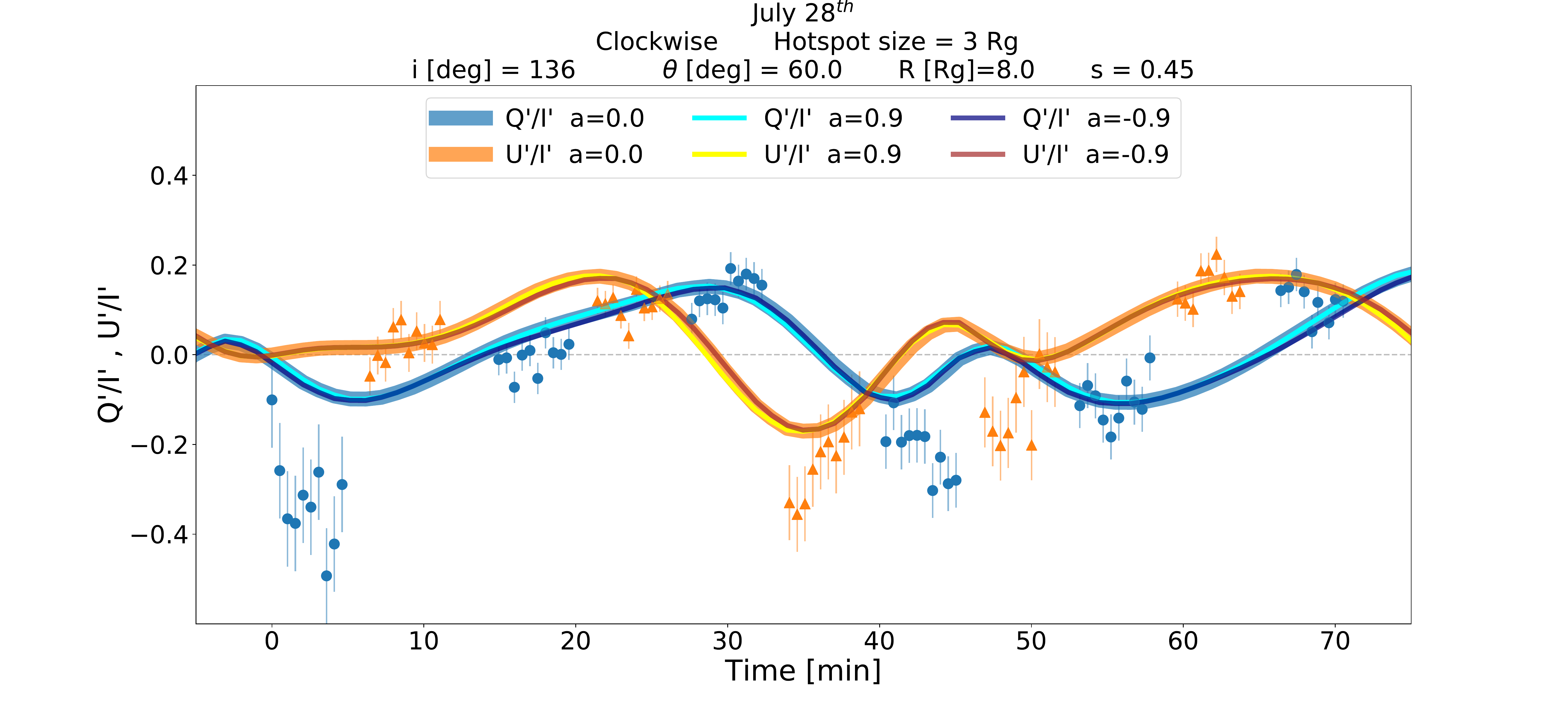}
\caption{Best fit model of the July $28^{\rm }$ flare calculated with three different values of dimensionless spin (a= $0.0,0.9,-0.9$). The reduced $\chi^2$ are reported in Table \ref{table:spin_comparison}. Changes in spin do not affect the curves significantly.}
\label{fig:spin_comparison}
\end{figure*}

We present the effects of spin in our calculations. Figure \ref{fig:spin_comparison} shows three models with the best fit parameters found for the July $28^{\rm }$ flare, at three different dimensionless spin values a=$0.0,0.9,-0.9$. The corresponding reduced $\chi^2$ values are reported in Table \ref{table:spin_comparison}. It can be seen that changes in spin do not alter the curves significantly and they can therefore be ignored.

\section{Scaling period effects}
\label{appendix:radii_comparison}

\begin{figure*}
\centering
\includegraphics[trim = 0cm 0 0cm 0cm, clip=true,width=0.8\textwidth]{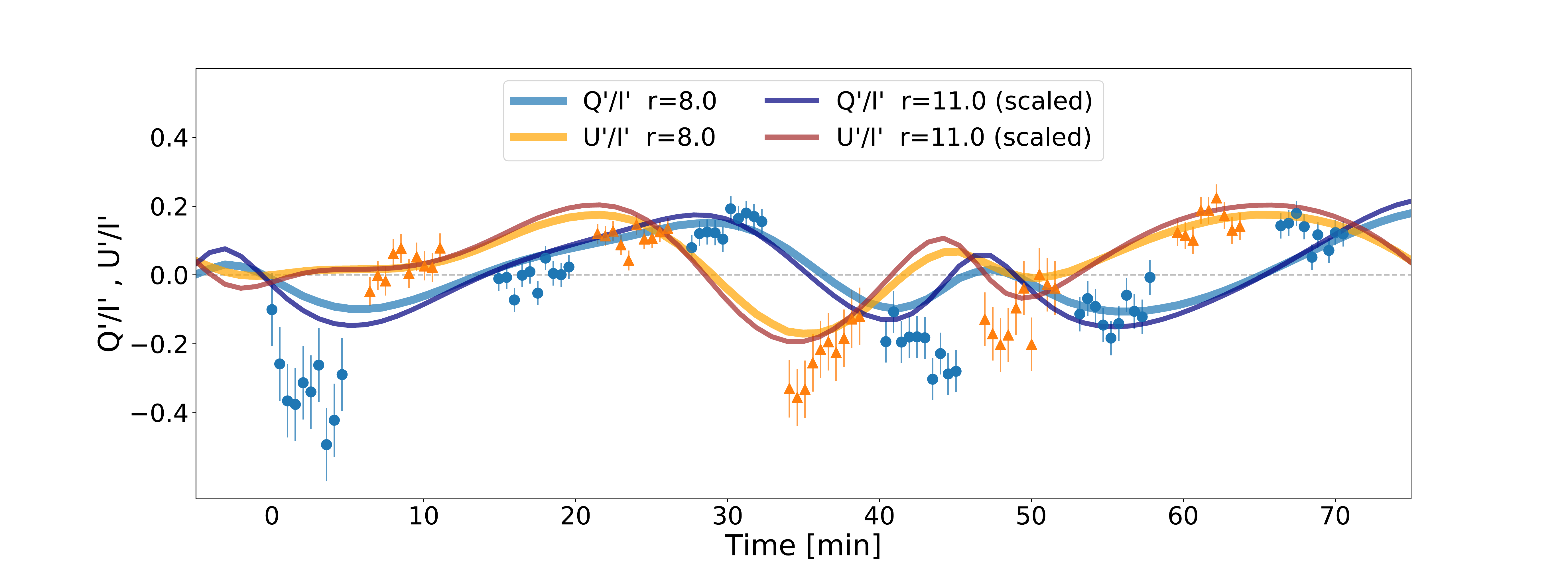}
\caption{Models calculated at $R=8\ R_g$ and at $R=11\ R_g$, the latter was scaled down to match the orbital period at $8\ R_g$. The rest of the parameters are those found for the best fit for the July $28^{\rm }$ flare. The reduced $\chi^2$ are reported in Table \ref{table:radii_comparison}. For better clarity, the $R=11\ R_g$ non-scaled model fit is not shown, but the $\chi^2$ is reported.
 }
\label{fig:radii_comparison}
\end{figure*}

We explore the effects of scaling the period of model curves. Figure \ref{fig:radii_comparison} shows the best fit model found for the July $28^{\rm }$ flare and one calculated at $R=11\ R_g$ scaled down to match the period at $8\ R_g$, with the rest of the parameters fixed to those of the best fit. The corresponding reduced $\chi^2$ values are reported in Table \ref{table:radii_comparison}. It can be seen that the curves show similar behaviours. Scaled models might have a better reduced $\chi^2$ than their non-scaled versions, but they are still not better than the best fit.

\begin{table}
\caption{Reduced $\chi^2$ of models calculated at $R=8\ R_g$ and at $R=11\ R_g$, the latter was scaled down to match the orbital period at $8\ R_g$. }

\label{table:radii_comparison}      
\centering                          
\begin{tabular}{c c c}        
\hline
 R  $[\rm{R_g}]$ & $a$ & $\chi^2$\\    
\hline                        
   8.0 & 0.0 & $3.104$ \\      
   11.0 (scaled) & 0.0 & $3.256$ \\      
   11.0 (not scaled) & 0.0 & $6.424$ \\      
   \hline                                
\end{tabular}
\end{table}

\section{Qualitative beam depolarisation}
\label{appendix:LP_discussion}
\begin{figure*}
\centering
\includegraphics[trim = 0cm 0cm 0cm 0cm, clip=true,width=0.75\textwidth]{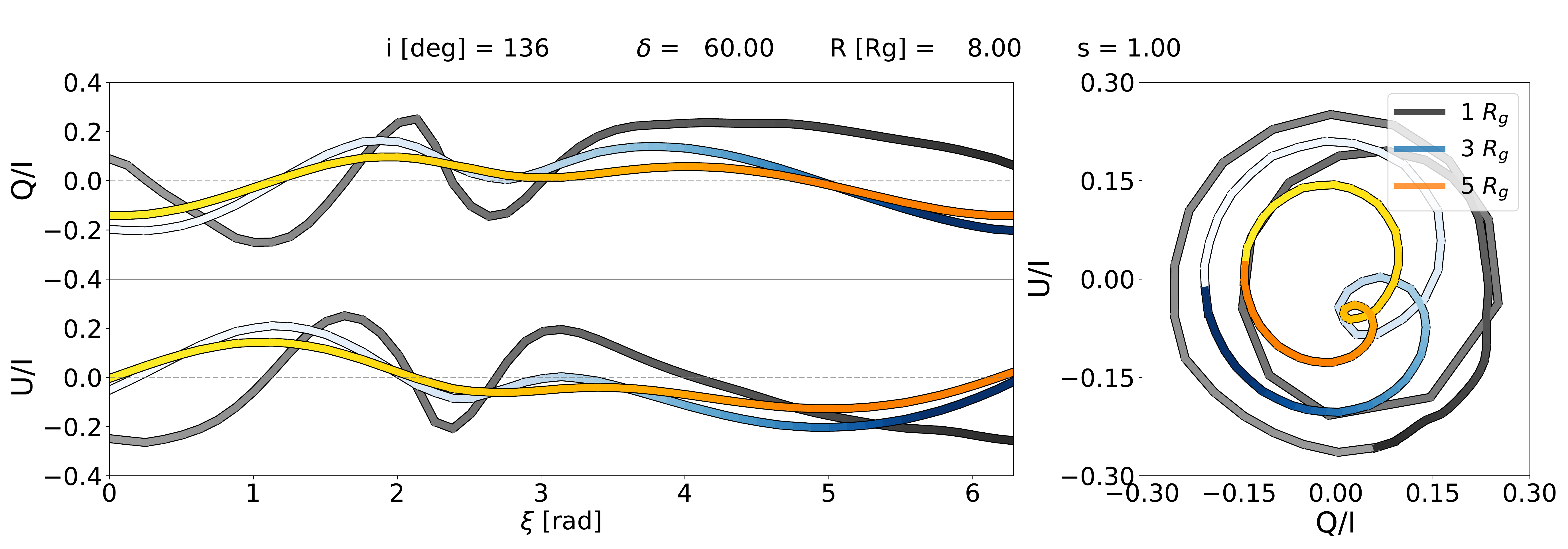}
\caption{Comparison of three numerical calculations with all identical parameters, except for $R_{\rm spot}$: $1$, $3$, and $5 \ \rm{R_g}$. As the hotspot size increases, the curve features are smoothed from beam depolarisation by sampling larger magnetic field regions and averaging out the different polarisation directions in time.}
\label{fig:size_comparison}
\end{figure*}

In the absence of other mechanisms, such as self-absorption or Faraday rotation and conversion, infrared emission from an orbiting hotspot is depolarised by beam depolarisation. 
Beam depolarisation works by capturing different contributions from polarisation (or magnetic field) structure and averaging them out.

More beam depolarisation occurs, the larger the emitting region that samples the underlying magnetic field is, or the more disordered the field itself is.
Given the simple magnetic field geometries considered in this work, disorder at small scales is non-existent. We discuss qualitatively the impact of emission size in the following.

As the hotspot goes around the black hole, it samples a wedge of angles in the azimuthal direction with an arc length of $R_{\rm spot}/R_0$.
Larger beam depolarisation occurs with the increase of this factor.  
Figure  \ref{fig:size_comparison} shows example curves of numerical calculations at a moderate inclination and magnetic field tilt, where only the hotspot size has been changed.
As expected, with increasing $\rm{R_{spot}}$ at a fixed orbital radius, not only does the amplitude of the polarised curves and $QU$ loops diminish (and with it, the linear polarisation fraction), but the features in them are smoothed out as well. 
Within the hotspot model, beam depolarisation can therefore be used to constrain the size of the emitting region as a function of the observed linear polarisation fraction.

\end{appendix}

\end{document}